\documentstyle[11pt,amssymb]{article}

\makeatletter
\@addtoreset{equation}{section}
\makeatother

\makeatletter
\def\citen#1{%
\edef\@tempa{\@ignspaftercomma,#1, \@end, }
\edef\@tempa{\expandafter\@ignendcommas\@tempa\@end}%
\if@filesw \immediate \write \@auxout {\string \citation {\@tempa}}\fi
\@tempcntb\m@ne \let\@h@ld\relax \def\@citea{}%
\@for \@citeb:=\@tempa\do {\@cmpresscites}%
\@h@ld}
\def\@ignspaftercomma#1, {\ifx\@end#1\@empty\else
   #1,\expandafter\@ignspaftercomma\fi}
\def\@ignendcommas,#1,\@end{#1}
\def\@cmpresscites{%
 \expandafter\let \expandafter\@B@citeB \csname b@\@citeb \endcsname
 \ifx\@B@citeB\relax 
    \@h@ld\@citea\@tempcntb\m@ne{\bf ?}%
    \@warning {Citation `\@citeb ' on page \thepage \space undefined}%
 \else
    \@tempcnta\@tempcntb \advance\@tempcnta\@ne
    \setbox\z@\hbox\bgroup 
    \ifnum0<0\@B@citeB \relax
       \egroup \@tempcntb\@B@citeB \relax
       \else \egroup \@tempcntb\m@ne \fi
    \ifnum\@tempcnta=\@tempcntb 
       \ifx\@h@ld\relax 
          \edef \@h@ld{\@citea\@B@citeB }%
       \else 
          \edef\@h@ld{\hbox{--}\penalty\@highpenalty
            \@B@citeB }%
       \fi
    \else   
       \@h@ld\@citea\@B@citeB
       \let\@h@ld\relax
 \fi\fi%
 \def\@citea{,\penalty\@highpenalty\hskip.13em plus.1em minus.1em}%
}
\def\@citex[#1]#2{\@cite{\citen{#2}}{#1}}%
\def\@cite#1#2{\leavevmode\unskip
  \ifnum\lastpenalty=\z@\penalty\@highpenalty\fi
  \ [{\multiply\@highpenalty 3 #1
      \if@tempswa,\penalty\@highpenalty\ #2\fi 
    }]\spacefactor\@m}
\makeatother

\let\a=\alpha   \let\d=\delta 
    
  \let\n=\nu

\let\la=\label  
   
\def\nn{\nonumber} \def\bd{\begin{document}} \def\ed{\end{document}}
\def\ds{\documentstyle} \let\fr=\frac \let\bl=\bigl \let\br=\bigr
\let\Br=\Bigr \let\Bl=\Bigl 
\let\bm=\bibitem
\let\na=\nabla
\let\pa=\partial \let\ov=\overline 
\newcommand{\be}{\begin{equation}} 
\newcommand{\ee}{\end{equation}} 
\def\ba{\begin{array}}
\def\ea{\end{array}}
\def\ft#1#2{{\textstyle{{\scriptstyle #1}\over {\scriptstyle #2}}}}
\def\fft#1#2{{#1 \over #2}}
\def\del{\partial}
\def\vp{\varphi}
\def\sst#1{{\scriptscriptstyle #1}}
\def\oneone{\rlap 1\mkern4mu{\rm l}}
\def\simequiv{\buildrel\sim\over=}
\def\td{\tilde}
\def\wtd{\widetilde}
\def\ie{{\it i.e.\ }}
\def\im{{\rm i}}
\def\dalemb#1#2{{\vbox{\hrule height .#2pt
        \hbox{\vrule width.#2pt height#1pt \kern#1pt
                \vrule width.#2pt}
        \hrule height.#2pt}}}
\def\square{\mathord{\dalemb{6.8}{7}\hbox{\hskip1pt}}}
\def\R{\rlap{\rm I}\mkern3mu{\rm R}}
\def\sR{\rlap{\hbox{$\scriptstyle\rm I$}}\mkern3mu{\hbox{$
\scriptstyle\rm R$}}}
\def\E{\rlap{\rm I}\mkern3mu{\rm E}}
\def\Z{\rlap{\sf Z}\mkern3mu{\sf Z}}
\def\0{{\sst{(0)}}}
\def\1{{\sst{(1)}}}
\def\2{{\sst{(2)}}}
\def\3{{\sst{(3)}}}
\def\4{{\sst{(4)}}}
\def\5{{\sst{(5)}}}
\def\6{{\sst{(6)}}}
\def\7{{\sst{(7)}}}
\def\8{{\sst{(8)}}}
\def\n{{\sst{(n)}}}
\def\v{{\cal V}}
\def\semi{\ltimes}

\newcommand{\ho}[1]{$\, ^{#1}$}
\newcommand{\hoch}[1]{$\, ^{#1}$}
\newcommand{\bea}{\begin{eqnarray}} 
\newcommand{\eea}{\end{eqnarray}} 
\newcommand{\ra}{\rightarrow}
\newcommand{\lra}{\longrightarrow}
\newcommand{\Lra}{\Leftrightarrow}
\newcommand{\ap}{\alpha^\prime}
\newcommand{\bp}{\tilde \beta^\prime}
\newcommand{\tr}{{\rm tr} }
\newcommand{\Tr}{{\rm Tr} } 
\newcommand{\NP}{Nucl. Phys. }

\newcommand{\tamphys}{\it Center for Theoretical Physics,
Texas A\&M University, College Station, Texas 77843}
\newcommand{\ens}{\it Laboratoire de Physique Th\'eorique de l'\'Ecole
Normale Sup\'erieure\hoch{2}\\
24 Rue Lhomond - 75231 Paris CEDEX 05}
\newcommand{\ic}{\it The Blackett Laboratory, Imperial College,\\
Prince Consort Road, London SW7 2BZ, UK}

\newcommand{\auth}{E. Cremmer\hoch{\dagger}, I.V. Lavrinenko\hoch{\ddagger}, 
H. L\"u\hoch{\dagger}, C.N. Pope\hoch{\ddagger1}, K.S. Stelle\hoch{\star}
and T.A. Tran\hoch{\ddagger}}

\begin{document}
\null
\vspace{-2cm}
\begin{flushright}
\hfill{CERN-TH/98-102}\\
\hfill{CTP TAMU-13/98}\\
\hfill{Imperial/TP/97-98/34}\\
\hfill{LPTENS-98/12}\\
\hfill{hep-th/9803259}\\
\hfill{March 1998}\\
\end{flushright}

\begin{center}
{\bf{\large Euclidean-signature Supergravities,
Dualities and Instantons}\hoch{\diamondsuit}}

\vspace{15pt}
\auth

\vspace{10pt}

{\hoch{\dagger}\ens}

\vspace{10pt}

{\hoch{\ddagger}\tamphys}

\vspace{10pt}
{\hoch{\star}\ic}

\vspace{15pt}

\underline{ABSTRACT}
\end{center}

     We study the Euclidean-signature supergravities that arise by
compactifying $D=11$ supergravity or type IIB supergravity on a torus
that includes the time direction.  We show that the usual T-duality
relation between type IIA and type IIB supergravities compactified on
a spatial circle no longer holds if the reduction is performed on the
time direction.  Thus there are two inequivalent Euclidean-signature
nine-dimensional maximal supergravities.  They become equivalent upon
further spatial compactification to $D=8$.  We also show that duality
symmetries of Euclidean-signature supergravities allow the harmonic
functions of any single-charge or multi-charge instanton to be
rescaled and shifted by constant factors.  Combined with the usual
diagonal dimensional reduction and oxidation procedures, this allows
us to use the duality symmetries to map any single-charge or
multi-charge $p$-brane soliton, or any intersection, into its
near-horizon regime. Similar transformations can also be made on
non-extremal $p$-branes.  We also study the structures of duality
multiplets of instanton and $(D-3)$-brane solutions.

{\vfill\leftline{}\vfill
\vskip  10pt
\footnoterule
{\footnotesize \hoch{\diamondsuit} Research supported in part by the
European Commission under TMR contract \vskip -12pt} \vskip 10pt
{\footnotesize \hoch{\phantom{\diamondsuit}} 
ERBFMRX-CT96-0045 \vskip -12pt} \vskip 14pt
{\footnotesize  \hoch{1} Research supported in part by DOE 
Grant DE-FG03-95ER40917 \vskip  -12pt} \vskip 14pt
{\footnotesize
        \hoch{2} Unit\'e Propre du Centre National de la Recherche
Scientifique, associ\'ee \`a l'\'Ecole Normale Sup\'erieure
\vskip -12pt} \vskip 10pt
{\footnotesize \hoch{\phantom{2}} et \`a l'Universit\'e de Paris-Sud
\vskip -12pt} \vskip 14pt}

\pagebreak
\setcounter{page}{1}

\section{Introduction}

     The study of dimensional reduction from eleven-dimensional
supergravity or type IIB supergravity is of great interest for a
variety of reasons.  In particular, the U-duality symmetries \cite{ht}
become more apparent in lower dimensions.  Not only are the
lower-dimensional supergravity theories of intrinsic interest in their
own right, but they also provide an organised way of studying classes
of solutions to the higher-dimensional equations of motion that
possess certain continuous symmetries.  This is particularly relevant
for the study of the various $p$-brane solitons that play such a
central r\^ole in some of the recent advances in understanding duality
symmetries in string theory and M-theory.  The most commonly
considered symmetries are translational symmetries along the spatial
directions in the $p$-brane world-volume.  Since the standard kind of
$p$-brane solution is Poincar\'e invariant on the world-volume, it
follows that the solution can be diagonally dimensionally reduced,
from a $p$-brane in $D+1$ dimensions to a $(p-1)$ brane in $D$
dimensions.  The consistency of the Kaluza-Klein dimensional reduction
procedure ensures that if the $p$-brane solves the $(D+1)$-dimensional
equations of motion, then the $(p-1)$-brane will solve the
$D$-dimensional equations of motion.  Thus the spatially dimensionally
reduced $D$-dimensional supergravities provide the arena within which
the dimensionally reduced, or ``wrapped,'' brane configurations can be
described.

     It is also, of course, the case that the standard $p$-brane
solutions are {\it static} (or {\it stationary} in the case of
rotating configurations).  Thus there is also an isometry in the time
direction, and so it is possible to interpret such configurations from
a lower-dimensional point of view as solutions of Euclidean-signature
supergravities.  These theories are obtained from the usual
Minkowskian-signature eleven-dimensional supergravity, or type IIB
supergravity, by performing a sequence of Kaluza-Klein reductions that
includes a reduction on the time direction.  Such Euclidean-signature
theories have been much less well studied than the
Minkowskian-signature ones (but see \cite{moore}) , and it is with
various aspects of the former that we shall be principally concerned
in this paper.

    Among our results, one of the more striking is that in nine
dimensions there are actually two inequivalent Euclidean-signature
maximal supergravities, one that is obtained from the reduction of
type IIA supergravity on the time direction, and the other that comes
from the reduction of type IIB supergravity on the time direction.
Usually, if a spatial reduction is performed, the two nine-dimensional
theories are equivalent, up to real field redefinitions.  This is the
field-theory precursor of the T-duality of the type IIA and type IIB
strings.  However, in the time reduction that we are considering here,
the two nine-dimensional theories are distinct, and cannot be related
to one another by any valid field redefinition.  It is only after a
further reduction of the two Euclidean-signature theories to $D=8$
that an equivalence emerges.

      One of the motivations for investigating Euclidean-signature
supergravities is to study the instanton states, which necessarily
live in Euclidean-signature space.  Unlike $p$-branes with $p\ge0$,
which are supported by higher-degree field strengths, and which form
linear representations under the U-duality group, the instantons are
supported by axionic scalars, which transform non-linearly under
U-duality.  The orbits of the higher $p$-branes in M-theory are much
better understood, and were obtained in \cite{fg,lpsorbit}.  In this
paper, we shall study the U-duality transformations of instanton
solutions, and also the orbits of their charges, which are the Noether
charges of the global symmetry group.

      Another of our results is concerned with the properties of the
instanton solutions that are the natural end-points of a sequence of
diagonal reductions of $p$-branes, when the reduction has encompassed
the entire world-volume including the time direction.  We show that
all instanton solutions, including multi-charge ones and even
non-extremal ones, have the property that they can be transformed,
using $SL(2,\R)$ global duality symmetries of the lower-dimensional
theories, into solutions where the harmonic functions characterising
the solutions are shifted and scaled by constants.  In particular, the
shifts can be chosen so as to remove the constant terms in the
harmonic functions altogether, with the result that for extremal
$p$-branes the entire solution is of the form that was previously
approached only asymptotically in the near-horizon limit.  The
solutions can then be oxidised back to higher dimensions, by retracing
the sequence of reduction steps.  They then describe $p$-branes again,
but now with similarly shifted harmonic functions.  Thus the
asymptotic structure of {\it any} extremal $p$-brane can be modified,
by such duality transformations, to have its near-horizon form.  In
the case of $p$-branes where the dilatons are finite on the horizon,
this means that the solutions are mapped into the AdS$\times$Sphere
form, where the supersymmetry is enhanced.  For non-extremal
$p$-branes, the structure of the outer horizon is governed by a
function that is not transformed under the $SL(2,\R)$ symmetry, and so
the effect of the modifications to the harmonic functions is more
complicated. A similar idea for transforming the asymptotic structure
of solutions was first proposed in \cite{hyun}, and developed in
\cite{bps}, using a different procedure in which a sequence of
T-duality and S-duality transformations were used to map the $p$-brane
to a wave, on which a general coordinate transformation was then
performed, followed by a retracing of the steps of the duality
transformations. This again has the effect of shifting and scaling the
harmonic functions.  More recently, another approach was given in
which the $SL(2,\R)$ duality of the Euclideanised type IIB theory was
used \cite{bb} instead of the general coordinate transformation on a
wave.  Our approach is simpler than either of these, since it just
involves diagonal reduction and oxidation, with an $SL(2,\R)$
transformation on the instanton at the bottom of the chain.
   
     The paper is organised as follows.  In section 2, we construct
the bosonic sectors of the Euclidean-signature maximal supergravities
that are obtained by dimensional reduction on a torus that includes
the time direction.  We also give an explicit demonstration that the
resulting lower-dimensional theories are insensitive to the order in
which the time and spatial reduction steps are performed.  In section
3, we discuss the global symmetries of the $D$-dimensional
Euclidean-signature theories, showing that they have the same
$E_{n(+n)}$ form as in the case of Minkowskian signature, where
$n=11-D$.  However, the denominator groups in the description of the
scalar cosets are no longer the maximal compact subgroups of
$E_{n(+n)}$, but instead certain non-compact forms of the previous
denominator groups, and we determine these for all $D$.  In section 4,
we consider the nine-dimensional Euclidean-signature theory obtained
by reducing type IIB supergravity on the time direction, and we show
that it is inequivalent to the nine-dimensional theory obtained by
reducing type IIA supergravity on the time direction.  In section 5,
we examine extremal instanton solutions in an $SL(2,\R)$-invariant
Euclidean-signature theory, discussing in detail how the symmetry acts
on the solutions.  We also consider non-extremal instantons, and show
that solutions exist only in an enlarged theory with at least a
$GL(2,\R)$ global symmetry.  We also study the effects of the global
symmetry transformations on the asymptotic structures of $p$-branes.
In section 6, we consider instantons in an $SL(3,\R)$-invariant
theory.  The action of the global symmetries on instanton solutions in
this case gives a better understanding of the general situation when a
number of different axions are capable of supporting the solution.  In
section 7, we consider $(D-3)$-brane solutions.  Although these are in
some sense the magnetic duals of the instantons, their structure is
very different for a variety of reasons, including the fact that they
live in Minkowskian-signature theories, and that their transverse
spaces are only two-dimensional.  The paper ends with a discussion in
section 8.

\section{Kaluza-Klein reduction on time}

    The standard categories of $p$-brane soliton solutions in
supergravities can be extended to the case of $p=-1$.  These
$(-1)$-branes have ``world-volumes'' of dimension $p+1=0$, and so all
the dimensions are occupied by the transverse space.  This means that
there is no longer any timelike dimension, and the solution is an
instanton in a purely Euclidean-signature space.  There are two ways
that such Euclidean-signature theories can arise.  The first is if we
take a standard supergravity theory in a $D$-dimensional spacetime,
and perform a Wick rotation of the time coordinate and reformulate the
theory in a $D$-dimensional space of Euclidean metric signature.  This
is a potentially problematic procedure; it might well be that the
original Lorentzian-signature supergravity involved the use of
fermions satisfying a Majorana condition, which can no longer be
covariantly imposed if the spacetime signature is altered.  Or, as in
the the case of the type IIB theory in $D=10$, the self-duality
constraint on the 5-form field strength cannot be imposed if the
spacetime is Euclideanised.

    A much more satisfactory situation obtains in cases where a
supergravity theory is dimensionally reduced on its time direction.
In such a case, the resulting lower-dimensional theory naturally
arises with a Euclidean-signature metric, and the consistency of the
reduction procedure guarantees that any Majorana or self-duality
constraints will be compatible with the Euclidean signature.  From the
point of view of the $p$-brane solutions, the instantons can be viewed
as the final stage of a sequence of diagonal dimensional reductions,
in which the time dimension of the ``world-volume'' of a 0-brane, or
static black hole, is dimensionally reduced in the final reduction step.

    In this paper, we shall principally focus our attention on
Euclidean-signature supergravities of this latter type, which are
obtained by dimensional reduction on the time coordinate.  In order to
study these theories in detail, it is useful to repeat an analysis
given in \cite{stainless}, for the single-step Kaluza-Klein reduction
of a the metric tensor and a generic gauge potential of degree
$(n-1)$, but where we now take the reduction to be on the time
coordinate.  Combined with the usual rules for spacelike reductions,
we can follow a route from $D=11$ or $D=10$ to any desired lower
dimension $D$, with the time reduction occurring at any desired stage in
the process.

\subsection{Bosonic Lagrangians}

     Let us suppose that we start with the Lagrangian
\be
{\cal L} = \hat e\, \hat R - \ft12\hat e\, (\del\hat \phi)^2 - 
\fft{1}{2n!}\hat e\,  e^{\hat a\hat \phi} \hat F_\n^2\ ,\label{hatlag}
\ee
in $(D+1)$ spacetime dimensions, where $\hat F_\n = d\hat A_{\sst(n-1)}$.  
We now perform a Kaluza-Klein reduction on the time coordinate, making the
ans\"atze
\bea
d\hat s^2 &=& e^{-2\alpha\varphi}\, ds^2 - e^{2(D-2)\alpha\varphi}\,
(dt + {\cal A}_\1)^2\ ,\nn\\
\hat A_{\sst(n-1)}(x,t) &=& A_{\sst(n-1)}(x) + A_{\sst(n-2)}(x) \wedge dt
\ ,\nn\\
\hat \phi(x,t) &=& \phi(x)\ ,\label{tkkred}
\eea 
where $\alpha=(2(D-1)(D-2))^{-1/2}$.  Substituting into (\ref{hatlag}), 
we obtain the reduced Lagrangian in $D$ spatial dimensions:
\bea
{\cal L} &=& e\, R -\ft12 e\, (\del \phi)^2 -\ft12 e\, (\del \varphi)^2
+\ft14 e\, e^{2(D-1)\alpha\varphi}\, {\cal F}_\2^2\nn\\
&&-\ft1{2\, n^!}\, e\, e^{2(n-1)\alpha\varphi +\hat a\phi}\, 
F_\n^2 +\ft1{2\, (n-1)!}\, e\, e^{-2(D-n)\a\varphi +\hat a\phi}\, 
F_{\sst(n-1)}^2 \ ,\label{delag}
\eea
where
\bea
F_\n &=& dA_{\sst(n-1)} -dA_{\sst (n-2)}\wedge {\cal A}_\1\ ,\nn\\
F_{\sst(n-1)} &=& dA_{\sst (n-2)}\ .\label{kkterms}
\eea
The reduced Lagrangian (\ref{delag}) differs from the usual one that
arises from a reduction on a spacelike coordinate in the signs of the
kinetic terms for ${\cal F}_\2$ and $F_{\sst(n-1)}$.  In this paper,
we call a field with the standard sign ``$-$'' for its kinetic
term a C-field (compact field) and a field with a ``$+$'' sign for its
kinetic term an NC-field (non-compact field).

    We are now in a position to present the general results for the
form of the $D$-dimensional maximal supergravity, where one of the
dimensional reduction steps may be on the timelike coordinate.  The
Kaluza-Klein ansatz for the reduction of the metric will be \cite{cjlp}
\be
ds_{11}^2 = e^{-\ft13\vec a\cdot\vec\phi}\, ds_{\sst D}^2 + \sum_i 
\varepsilon_i\, e^{2\vec\gamma_i\cdot\vec\phi}\, 
(dz^i + {\cal A}_\1^i + {\cal A}^i_{\0 j}\, dz^j)^2\ ,\label{dmetric}
\ee
where $\vec\gamma_i=-\ft16 \vec a+\ft12 \vec b_i$, with $\vec a$ and
$\vec b_i$ being the dilaton vectors for $F_\4$ and ${\cal F}_\2^i$,
as defined in \cite{lpsol,cjlp}, and discussed below.  
The constants $\varepsilon_i$ are
$+1$ for spacelike coordinate reduction steps, and $-1$ for a timelike
step.  In the notation of \cite{lpsol,cjlp}, the $D$-dimensional 
Lagrangian will therefore be
\bea
{\cal L} &=& eR -\ft12 e\, (\del\vec\phi)^2 -\ft1{48}e\, e^{\vec a\cdot
\vec\phi}\, F_\4^2 -\ft{1}{12} e\sum_i \varepsilon_i\, 
e^{\vec a_i\cdot \vec\phi}\, (F_{\3i})^2 \nn\\
&&-\ft14 e\, \sum_{i<j} \varepsilon_i\, \varepsilon_j\, 
e^{\vec a_{ij}\cdot \vec\phi}\, (F_{\2ij})^2
-\ft14e\, \sum_i \varepsilon_i\, 
e^{\vec b_i\cdot \vec\phi}\, ({\cal F}_\2^i)^2\label{dgenlag}\\
&&-\ft12 e\, \sum_{i<j<k}\varepsilon_i\, \varepsilon_j\, \varepsilon_k\,
e^{\vec a_{ijk} \cdot\vec \phi}\,
(F_{\1ijk})^2
 -\ft12e\, \sum_{i<j}\varepsilon_i\, \varepsilon_j\,  
e^{\vec b_{ij}\cdot \vec\phi}\, ({\cal F}_\1^i{}_j)^2 + 
{\cal L}_{\sst{FFA}}\ ,\nn
\eea
where the dilaton vectors $\vec a$, $\vec a_i$, $\vec a_{ij}$, $\vec
a_{ijk}$, $\vec b_i$, $\vec b_{ij}$ are constants that characterise
the couplings of the dilatonic scalars $\vec \phi$ to the various
gauge fields.  Their detailed expressions, together with the
Kaluza-Klein modifications in the various field strengths, are given
in \cite{lpsol,cjlp}.  ${\cal L}_{\sst{FFA}}$ is the Wess-Zumino term,
whose detailed expression can also be found in \cite{lpsol,cjlp};
since this is written without the use of the metric, it is the same as
in the usual purely spatial reductions.  Note that the indices
$i,j\ldots$ range over the internal compactified dimensions, starting
with $i=1$ for the reduction step from $D=11$ to $D=10$.  Thus in the
situation where the $N$'th reduction step is on the time coordinate,
the signs of certain of the kinetic terms will be reversed, relative
to the Minkowskian-signature case, as is evident in (\ref{dgenlag}).
Specifically, we can see that these are the kinetic terms for all
field strengths that carry an internal index equal to the value $N$.

\subsection{Commutativity of time and space reductions}

     We have seen that the $D$-dimensional Euclidean-signature theory
that is obtained by compactifying on the time coordinate at the step
$i$, and on spatial coordinates at all other steps, is given by eqn
(\ref{dgenlag}), and that the signs of all the kinetic terms whose
fields involve the index value ``$i$'' are reversed in comparison to
the ``standard'' signs of the Minkowskian-signature theory.  It is not
{\it a priori} obvious that the Euclidean-signature theory in $D$
dimensions is the same regardless of which step is chosen for the time
reduction.  In this section, we present a proof which demonstrates
that all such $D$-dimensional theories are in fact related by valid
field redefinitions.  Specifically, we shall concentrate on the scalar
subsectors of the $D$-dimensional theories.  The proof extends to the
full bosonic Lagrangians.

    To do this, we note that the Lagrangian (\ref{dgenlag}) was
obtained directly from the dimensional reduction of $D=11$
supergravity without any dualisations.  This Lagrangian has a
$GL(11-D,\R)\semi R^q$ global symmetry, where $q=\ft16 (11-D) (10-D)
(9-D)$ \cite{lptdual,cjlp}.  (The $E_{n(+n)}$ global symmetry
\cite{cj2} is achieved by performing all Hodge dualisations that turn
higher-degree fields into lower-degree ones.)  The $GL(11-D,\R)$
symmetry is generated by the full set of dilatonic and axionic scalars
coming from the metric; the higher-degree fields and the remaining
axionic scalars coming from the reduction of the 3-form potential in
$D=11$ form linear representations under $GL(11-D,\R)$.  We recall
from \cite{cjlp} that the scalar coset manifold for $SL(11-D,\R)$ can
be parameterised in the Borel gauge as
\be
\v = e^{\ft12\vec\phi\cdot\vec H}\, h\ ,\label{cosetv}
\ee
where $\vec H$ is the set of Cartan generators for the $SL(11-D,\R)$ global
symmetry group, $\vec \phi$ are the dilatons, and $h$ is
a parameterisation of the exponential of the positive-root algebra of
$SL(11-D,\R)$, with the axionic fields ${\cal A}^i_{\0 j}$ as the
parameters, {\it i.e.}
\be
h=\prod_{i<j} e^{ {\cal A}^i_{\0 j} E_i{}^j}\ ,
\ee
with the terms in the product arranged in anti-lexical order, namely
\be
(i,j) = \cdots (3,4), (2,4), (1,4), (2,3), (1,3), (1,2)\ .
\ee
The scalar Lagrangian for the $SL(11-D,\R)$ part is given by
\be
{\cal L} = \ft14 e\,  {\rm tr}(\del_\mu {\cal M}^{-1}\, 
\del^\mu{\cal M})\ ,\label{cosetlag}
\ee
where the matrix ${\cal M}$ is defined by
\be
{\cal M} = \v^T\, \eta\, \v\ ,\label{mmatrix}
\ee
and $\eta$ is a metric tensor.  In the usual case where one compactifies
$D=11$ supergravity on a set of spatial directions, $\eta$ is just the
identity.  If instead the $i$'th compactification coordinate is the time
coordinate, the results of section 2.1 show that the metric $\eta$ will have
the form $\eta=\eta(i)$, where
\be
\eta(i)\equiv {\rm diag}(1,\cdots, 1, -1, 1\cdots, 1)\ , \label{etai}
\ee
and the $-1$ occurs at the $i$'th position.  Note that as we shall
shown in section 3, the Lagrangian (\ref{cosetlag}) describes the
coset of $SL(11-D,\R)/O(10-D,1)$ when the time is one of the internal
coordinates, rather than $SL(11-D,\R)/O(11-D)$ when the internal
directions are all spatial.

   We wish to show that the $D$-dimensional Euclidean-signature theories
where the time compactification occurs at the $i$'th step are
equivalent, up to field redefinitions, for all $i$.  We may show this
in the following way.  The $i$'th theory is characterised completely by
the fact that the matrix ${\cal M}$ in (\ref{mmatrix}) is constructed
using $\eta=\eta(i)$, where $\eta(i)$ is defined in (\ref{etai}).  In
order to show that the theories for all $i$ are equivalent, we need to
show that there exist field redefinitions that relate them all.  To do
this, consider the $i$'th theory, and then make the following field
redefinition 
\be
\vec\phi\longrightarrow \vec\phi' = \vec\phi + \fft{i\pi}{2}\, 
\vec b_{ij}\ .\label{phitrans}
\ee
It is evident from (\ref{cosetv}) that this will transform the matrix
${\cal M}(i)=\v^T\, \eta(i)\, \v$, defined in (\ref{mmatrix}), according
to
\be
{\cal M}(i) \longrightarrow {\cal M}'(i) = 
      \v^T\, e^{\fft{i\pi}{4}\vec b_{ij}\cdot\vec H}\, 
       \eta(i)\,  e^{\fft{i\pi}{4}\vec b_{ij}\cdot\vec H}\,\v\ .
\ee
In fact we may take $\vec H$ and $\eta(i)$ to be diagonal, and so we
simply have 
\be
{\cal M}'(i) =\v^T\, \eta'(i)\, \v\ .\label{imatrix}
\ee
where 
\be
\eta'(i) = e^{\fft{i\pi}{2}\vec b_{ij}\cdot\vec H}\, \eta(i)\ .
\label{etaprime}
\ee

    To evaluate the expression $e^{\fft{i\pi}{2}\vec b_{ij}\cdot\vec H}$,
we may make use of the fact that the positive-root generators $E_i{}^j$
associated with the $SL(11-D,\R)$ subalgebra of the $E_{n(+n)}$ global
symmetry algebra satisfy the commutation relations \cite{cjlp}
\be
{[} \vec H, E_i{}^j {]} = \vec b_{ij}\, E_i{}^j\ .
\ee
Furthermore, the dilaton vectors $\vec b_{ij}$ have the property that
$\vec b_{ij}\cdot\vec b_{k\ell}$ is equal to 0 if $i$ and $j$ are
different from $k$ and $\ell$, while it equals $4$ if $i=k$ and
$j=\ell$, and $\pm2$ if there is just one index in common between $i$,
$j$, and $k$, $\ell$.  Since $E_i{}^j$ can be represented by the matrix
consisting of zeroes everywhere except for a ``1'' at the $i$'th row and
$j$'th column, we can deduce that the diagonal matrix $\vec
b_{ij}\cdot \vec H$ has entries equal to 2 mod 4 at the $i$'th and $j$'th
positions, and entries equal to 0 mod 4 at all other positions, on the
diagonal.  Thus we have that
\be
e^{\fft{i\pi}{2}\vec b_{ij}\cdot\vec H} = {\rm diag}\,
(1,\ldots1,-1,1\ldots,1,-1,1,\ldots 1)\ ,
\ee
where the $-1$ entries are at positions $i$ and $j$.  We see that
the metric $\eta'(i)$ defined in (\ref{etaprime}) is therefore simply
given by
\be
\eta'(i) = \eta(j)\ .
\ee
The conclusion from this is that if we start from the $D$-dimensional
Euclidean-signature Lagrangian in which the time reduction was
performed at step $i$, and make the field redefinition
(\ref{phitrans}), we end up with the Lagrangian that would be obtained
by making the time reduction instead at step $j$.

     So far we have concentrated on the scalars coming from the
metric, which generate the $SL(11-D,\R)$ global symmetry.  The field
redefinition (\ref{phitrans}) also provides proper sign changes for
the kinetic terms of the rest of the scalars and the higher forms as
well.  This can be seen from the fact that the dot products of the
dilaton vectors $(\vec a, \vec a_i, \vec a_{ij}, \vec a_{ijk}, \vec
b_i)$ for all the other fields with the dilaton vector $\vec b_{\ell
m}$ gives either $\pm 2$ if there is one common index, or 0 or 4 if
there is either no common index or two common indices.  Note that the
field redefinition (\ref{phitrans}) does not alter the signs of the
kinetic terms for all dilatonc scalars $\vec\phi$.  It has the effect
of reshuffling the signs of the kinetic terms of certain axions, and
higher-form potentials.  However, the total number of C-fields for
each degree (and hence also that of the NC-fields) is preserved under
this field redefinition.

     One might wonder about the validity of this construction, in view
of the fact that the field redefinition (\ref{phitrans}) involves
making an imaginary constant shift of the dilatons.  It is indeed true
that in general complex field redefinitions on real fields are not
permissible as a way of demonstrating the equivalence of ostensibly
different theories.  However, the crucial point here is that the
scalar manifolds in question are coset spaces, and provided that the
reality of the coset matrices ${\cal M}$ that are used in the
construction of the Lagrangians (\ref{cosetlag}) is maintained, then
the redefined $\vec\phi$ fields, even though subjected to imaginary
shifts, still provide a valid parameterisation, and the imaginary
parts have no physically-observable consequences. And indeed, we have
seen that the redefinition (\ref{phitrans}) simply has the effect of
replacing the real metric $\eta(i)$ in (\ref{imatrix}) by the real
metric $\eta(j)$, thus making manifest the continued reality of the
redefined matrix ${\cal M}'(i)$.

     In fact the transformation relating the step-$i$ and the step-$j$
Lagrangians can be viewed abstractly as a transformation between the
coset matrices ${\cal M}(i)$ and ${\cal M}(j)$, rather than a
transformation implemented explicitly on the coset coordinates
$\vec\phi$ and $\chi_a$.  In general, we can allow any transformation
of the form
\be
{\cal M}(i) \longrightarrow {\cal M}(j) = \Lambda\, {\cal M}(i)\,
\Lambda^{-1}\ ,\label{gentrans}
\ee
where $\Lambda$ is in the $E_{n(+n)}$ numerator group, since the
$\Lambda$ factors will cancel out in the Lagrangian (\ref{cosetlag}).
If $\Lambda$ is taken to be the identity, then this transformation
happens to be implementable in the form (\ref{phitrans}), in terms of
the parameterisation of $\v$ given by (\ref{cosetv}).  For other
parameterisations, or for other choices of $\Lambda$, the specific
form that the transformation (\ref{gentrans}) induces on the
coordinates of the coset will be different, and can, for example, be
arranged to be real, at least in some coordinate patch.  We give an
example of this later, in the case of an $SL(3,\R)/O(2,1)$ coset.  It
should be emphasised, however, that the equivalence of the two
Lagrangians is proved once the existence of a real transformation
between their respective coset matrices ${\cal M}$ is established.
Exhibiting explicit real, rather than complex, coset-coordinate
relations that implement this real transformation between the coset
matrices may be desirable for some purposes, but it is an inessential
part of the proof of equivalence of the Lagrangians.

    It is worth mentioning also that the transformation
(\ref{phitrans}) not only preserves the reality of the coset matrices
${\cal M}$, but it also preserves the reality of the original
eleven-dimensional metric.  This is an important point, since the
various fields in the $D$-dimensional theory, including the dilatons
$\vec\phi$, all originate from real fields in eleven dimensions.  To
be specific, the Kaluza-Klein ansatz for the $D$-dimensional metric
that was used in obtaining (\ref{dgenlag}) is given by
(\ref{dmetric}).  We now observe from \cite{lpsol} that $\vec
a\cdot\vec b_{ij}=0$, and $\vec b_k\cdot \vec b_{ij}=2\delta_{ik}
-2\delta_{jk}$.  Consequently, the effect of performing the field
redefinition $\vec\phi\rightarrow \vec\phi + \ft{i\pi}{2}\vec b_{ij}$
is to leave the entire eleven-dimensional metric (\ref{dmetric})
unchanged, except for the replacements
\be
\varepsilon_i\longrightarrow -\varepsilon_i\ ,\qquad
\varepsilon_j\longrightarrow -\varepsilon_j\ ,\qquad
\varepsilon_k\longrightarrow \varepsilon_k\ , \qquad k\ne i\ , k\ne j\ .
\ee
In other words, the effect of the transformation is precisely to
interchange which of $i$ and $j$ is the compactified time-like
direction.  The fact that the eleven-dimensional metric remains real
under the transformation (\ref{phitrans}) re-emphasises the fact that
it is not the $\vec\phi$ fields themselves that are physically
meaningful, but only the various exponentials of them that occur in
the metric and the $D$-dimensional Lagrangian.

     Having established that the field redefinition (\ref{phitrans})
maps the $D$-dimensional theory obtained by reducing on $t$ at the
$i$'th step to the theory obtained by instead reducing on $t$ at the
$j$'th step, it is evident that by choosing all possible dilaton
vectors $\vec b_{ij}$ in (\ref{phitrans}), we can establish the
equivalence of all the $D$-dimensional Euclidean-signature theories
obtained by this method.  In other words, the order in which the time
and space reductions are performed is immaterial.

          So far we have considered the bosonic Lagrangians obtained
from dimensional reduction of the $D=11$ Lagrangian without performing
any dualisation of the fields.  In order for the theories to have the
$E_{n(+n)}$ global symmetry groups of the maximal supergravities, it
is necessary to Hodge dualise all field strengths with degrees $> D/2$
to give fields of lesser degrees.  It is well known that this
dualisation procedure and Kaluza-Klein reduction on a torus commute.
In fact, dualisation commutes also with the reduction on the time
coordinate.  Let us illustrate this by a simple example.  Consider a
field strength $\hat F_\n$ in $(D+1)$-dimensional spacetime.  After a
dimensional reduction on the time direction, this field strength gives
rise to a compact field $F_\n$ and a non-compact field
$F_{\sst(n-1)}$.  Here we are calling fields with the ``standard''
sign for their kinetic terms compact fields, while those with the
non-standard sign are called non-compact fields.  Now let us dualise
the $\hat F_\n$ field strength to $\hat F_{\sst(D+1-n)}$ in $(D+1)$
dimensions.  After dimensional reduction on the time direction, this
field gives rise to a compact field $F_{\sst(D+1-n)}$ and a
non-compact field $F_{\sst(D-n)}$.  Now, Hodge dualisation in a
Euclidean-signature space always has the effect of changing the sign
of the kinetic term for any field of any degree, whilst dualisation in
a Minkowskian-signature spacetime always leaves the sign unaltered.
(Note that this dualisation property implies in particular that the
signs of the kinetic terms for $n$-form field strengths in a
Euclidean-signature space of even dimension $D=2n$ can be reversed at
will by dualisation.)  Thus in a $D$-dimensional Euclidean-signature
theory, the compact field strengths $F_\n$ and $F_{\sst(D+1-n)}$ are
dual to the non-compact fields $F_{\sst(D-n)}$ and $F_{\sst(n-1)}$
respectively.

\section{Cosets in Euclidean-signature spaces}

          In section 2, we obtained the bosonic Lagrangians for all
the maximal supergravities in Euclidean-signature spaces that come
from the dimensional reduction of eleven-dimensional supergravity with
time as one of the internal dimensions.  We have observed that the
Lagrangians are similar to those in Minkowskian-signature spacetimes,
except that the signs of the kinetic terms for certain fields are
reversed.  In this section, we show that the sign changes in these
kinetic terms do not alter the fact that these theories have
$E_{n(+n)}$ global symmetries, just as in the Minkowskian-signature
spacetimes. However, the denominator group $H$ of the coset
$E_{n(+n)}/H$ is no longer the maximal compact subgroup of
$E_{n(+n)}$.  It becomes instead a certain non-compact form of the
previous maximal compact subgroup.  In this section, we
shall determine these denominator groups for $D\ge 3$.

\subsection{An $SL(2,\R)$ example}

         Let us first examine the simplest non-trivial example, namely
the $SL(2,\R)$ system.  The nine-dimensional scalar Lagrangian in a
Euclidean-signature space is given by
\be
e^{-1} {\cal L} = -\ft12 (\del \phi)^2 + \ft12 e^{2\phi}\, (\del\chi)^2\ ,
\label{sl2rlag1}
\ee
together with a decoupled $O(1,1)$-invariant term
$-\ft12(\del\varphi)^2$ that does not concern us here.  This is to be
contrasted with the Minkowskian-signature Lagrangian
\be
e^{-1} {\cal L} = -\ft12 (\del \phi)^2 - \ft12 e^{2\phi}\, (\del\chi)^2\ ,
\label{sl2rlagmink}
\ee
Note that the axionic scalar $\chi$ in (\ref{sl2rlag1}) comes from the
dimensional reduction of the R-R vector in the $D=10$ type IIA theory,
and hence the sign of its kinetic term is reversed in the
Euclidean-signature space. (The situation is different in the time
reduction of the type IIB theory, which we shall discuss in section
4).  We now show that the Lagrangian (\ref{sl2rlag1}) is described by
the coset $SL(2,\R)/O(1,1)$.  To see this, we note that the Lagrangian
can be parameterised by the Borel subgroup of $SL(2,\R)$. Following
\cite{cjlp}, we can parameterise an $SL(2,\R)/O(1,1)$ coset
representative $\v$, in the Borel gauge, as
\be
\v = e^{\ft12\phi H}e^{\chi E_{+}}\,\,\,  =\,\,\, 
\pmatrix{e^{\ft12\phi} &\chi e^{\ft12\phi}\cr
                  0        & e^{-\ft12\phi} }\ ,\label{sl2rv}
\ee
where $H$ and $E_{+}$ are the Cartan and positive-root generators of
$SL(2,\R)$.  The Lagrangian (\ref{sl2rlag1}) can then be expressed as
\be
e^{-1} {\cal L} = \ft14 \tr (\del_\mu {\cal M}^{-1} \del^\mu {\cal
M})\ ,\label{gencosetlag}
\ee
where ${\cal M}$ is given by
\be
{\cal M} = \v^{\rm T} \eta \v=
\pmatrix{e^{\phi} & \chi e^{\phi} \cr
         \chi e^{\phi} & \chi^2 e^{\phi} -e^{-\phi}} \ ,\qquad 
\eta = {\rm diag}(1, -1) \ . \label{matrix}
\ee
Note that ${\rm Det}({\cal M}) = {\rm Det}(\eta) =-1$, so ${\cal M}$
is no longer an $SL(2,\R)$ matrix.  (In the case of the usual
$SL(2,\R)$ coset, where $\chi$ has the standard sign for its kinetic
term, $\eta$ would be ${\rm diag}(1,1)$, and hence ${\cal M}$ would be
an $SL(2,\R)$ matrix.)

      The global $SL(2,\R)$ transformations on the scalar fields can
be implemented by acting on the right of $\v$ with a constant
$SL(2,\R)$ matrix $\Lambda$, and on the left with a field-dependent
compensating $O(1,1)$ transformation ${\cal O}$, whose job is to
restore the transformed $\v$ to the Borel gauge:
\be
\v \longrightarrow \v'= {\cal O} \v \Lambda\ .\label{gencosettrans}
\ee
It is manifest that provided ${\cal O}$ satisfies ${\cal O}^{\rm T}
\eta {\cal O}=\eta$, this will leave the Lagrangian (\ref{gencosetlag})
invariant for any global $SL(2,\R)$ transformation.  Note that if the
axionic field $\chi$ had had the standard sign for its kinetic term,
as in the Minkowskian-signature Lagrangian (\ref{sl2rlagmink}), then
we would instead have $\eta={\rm diag}(1,1)$, and so ${\cal O}$ would
be an element of the compact group $O(2)$, implying that the coset
would be $SL(2,\R)/O(2)$.  In our Euclidean-signature Lagrangian
(\ref{sl2rlag1}), however, the opposite sign for the kinetic term for
$\chi$ implies that $\eta={\rm diag}(1,-1)$, and hence ${\cal O}$ is
an element of the non-compact group $O(1,1)$. Thus the Lagrangian
(\ref{sl2rlag1}) is described by the coset $SL(2,\R)/O(1,1)$.

        If we now introduce the pseudo-imaginary unit $j$, with $j^2=1$ 
and $\bar j =-j$, the fields $\chi$ and $\phi$ in this $SL(2,\R)/O(1,1)$
system can be grouped together as the double-number valued field $\tau =
\chi + j e^{-\phi}$.  The $SL(2,\R)$ global symmetry transformations can
then be expressed as the fractional linear transformation \cite{ggp}
\be
\tau \longrightarrow \tau' =\fft{a \tau + b}{c\tau +d}\ ,\label{sl2rtr}
\ee
where $ad - bc =1$.  In the more usual
Minkowskian-signature $SL(2,\R)/O(2)$ system, $j$ would be replaced by
the unit imaginary number $i$.

         In section 2.2, we showed that the signs of the kinetic terms
of certain of the scalar fields can be altered by making the field
redefinition (\ref{phitrans}), and by this means we established that
the processes of making dimensional reductions on time and space
coordinates commute.  In this $SL(2,\R)$ example, the difference
between the Lagrangians (\ref{sl2rlagmink}) and (\ref{sl2rlag1}) for
the cosets $SL(2,\R)/O(2)$ and $SL(2,\R)/O(1,1)$ is that the sign of
the kinetic term for the $\chi$ field is negative in the former case,
and positive in the latter.  This sign reversal could be achieved by
sending $\phi \rightarrow \phi + \ft{i}2 \pi$.  One might naively
deduce from this field redefinition that the cosets $SL(2,\R)/O(2)$
and $SL(2,\R)/O(1,1)$ were equivalent, a conclusion that is actually
false.  The reason for this is that under the redefinition $\phi
\rightarrow \phi+ \ft{i}2\pi$ the ${\cal M}$ matrix, which
parameterises the points in the scalar manifold, does not remain real,
unlike the situation in the cases we described in section 2.2. In
particular the string coupling constant $g=e^{-\phi}$ would become
imaginary.  In fact, for this reason, the field redefinition precisely
establishes the {\it inequivalence} of the two theories.  Furthermore,
when higher-degree field strengths $F_{\n}=dA_{\sst(n-1)}$ are
included in the Lagrangian, they couple to the scalars through terms
of the form of $F_{\n}^{\rm T} {\cal M} F_\n$.  By causing the matrix
${\cal M}$ to become complex, the field redefinition $\phi \rightarrow
\phi+ \ft{i}2\pi$ would also have the effect of making the Lagrangian
complex.  This emphasises the distinction between the valid complex
transformations of the kind we used in section 2.2 to show that two
ostensibly different Lagrangians are actually equivalent, and more
general kinds of complex transformation that change the structure of
the theory. Note that the analogue of the transformation
(\ref{phitrans}) in this $D=9$ example is $\phi \rightarrow \phi +
i\pi$, which does not change the sign of the kinetic term for $\chi$.

      The field redefinition (\ref{phitrans}) does have the effect of
making $\v$ become complex, but it leaves ${\cal M}$ real.  Of course
this field redefinition is not a symmetry of the theory, since it
changes the form of the Lagrangian.  In fact even transformations
under the global symmetries of the theory can also have the effect of
causing $\v$ to become complex, while again leaving ${\cal M}$ real.
The reality of ${\cal M}$ is guaranteed by the form of the global
transformation, namely ${\cal M} \rightarrow \Lambda^{\rm T} {\cal M}
\Lambda$, where $\Lambda$ is a real-valued matrix in the global
symmetry group $G$.  In a Euclidean-signature space ${\cal M}$ is not
positive definite, and hence $\v$, which can be viewed as a
square-root of ${\cal M}$, can be complex.  In a Minkowskian-signature
spacetime, by contrast, ${\cal M}$ {\it is} positive definite and so
$\v$ itself remains real under the global transformations.

\subsection{Cosets for maximal supergravities in Euclidean-signature
spaces} 

        The above demonstration can easily be generalised to lower
dimensions $D$, where the global symmetry groups are $E_{n(+n)}$ with
$n=11-D$.  The structure of ${\cal M}$ and (\ref{matrix}), and the
transformations (\ref{gencosettrans}), imply that only the denominator
local compensating group elements ${\cal O}$ will ``see'' the $\eta$
matrix, whilst the global group elements $\Lambda$ will be unaffected
by the signature change of $\eta$.  This shows that changing $\eta$
will not affect the global symmetry group, but it will change the
local denominator group $H$.

         To determine $H$, one can start by counting the numbers of
scalars that have standard or non-standard signs for their kinetic
terms.  As mentioned in section 2, we shall call the fields that have
standard-sign kinetic terms C-fields (compact fields), and those that
have the non-standard sign NC-fields (non-compact fields).  It is easy
to verify that the number of NC-scalars in the coset $G/H$ is the same
as the number of NC-generators (non-compact generators) in $H$. For
example, if there are no NC-scalars at all in the coset, as is the
case for the standard maximal supergravities in Minkowskian-signature
spacetime, then $H$ has no NC-generators, and hence it will be the
maximal compact subgroup of $G$.  In the above $SL(2,\R)$ example, we
have one NC-scalar, and hence one NC-generator in the denominator
group, implying that $H=O(1,1)$. We list in Table 1 all the scalars in
all the $D\ge3$ supergravities in Euclidean-signature spaces that come
from the dimensional reduction of eleven-dimensional supergravity
(including the scalars that are dualisations of all $(D-2)$-form
potentials).

\bigskip\bigskip

\begin{center}
\begin{tabular}{|c|c|c|c|c|c|}\hline
 & NC-scalars & C-scalars & 
$\pmatrix{\hbox{Total scalars}\cr =Dim(G/H)}$ & $Dim(G)$ & 
$Dim(H)$\\ \hline
$D=10$ &0   &1   &1    &1   & 0 \\   \hline
$D=9$  &1   &2   &3    &4   & 1 \\   \hline
$D=8$  &3   &4   &7    &11  & 4 \\   \hline
$D=7$  &6   &8   &14   &24  &10 \\   \hline
$D=6$  &10  &15  &25   &45  &20 \\   \hline
$D=5$  &16  &26  &42   &78  &36 \\   \hline
$D=4$  &27  &43  &70   &133 &63 \\   \hline
$D=3$  &56  &72  &128  &248 &120\\   \hline
\end{tabular}
\end{center}

\bigskip

\centerline{Table 1: Scalars in Euclidean-signature maximal supergravities}

\bigskip\bigskip

        It is now straightforward to determine the local denominator
groups $H$. For example, in $D=7$ we have that the dimension of $H$ is
10, comprising 6 NC-generators and 4 C generators, so we have
$H=O(3,2)$.  In Table 2, we summarise the cosets for all the
Euclidean-signature maximal supergravities in $D\ge 3$ (these results
can be found also in \cite{stelle,hj}).
\bigskip\bigskip

\begin{center}
\begin{tabular}{|c|c|c|}\hline
 & Minkowskian & Euclidean \\ \hline
$D=10$ & O(1,1) & O(1,1) \\ \hline
$D=9$ & $\fft{GL(2,\sR)}{O(2)}$  &  $\fft{GL(2,\sR)}{O(1,1)}$ \\ \hline
$D=8$ & $\fft{SL(3,\sR)\times SL(2,\sR)}{O(3)\times O(2)} $ &
        $\fft{SL(3,\sR)\times SL(2,\sR)}{O(2,1)\times O(1,1)}$ \\ \hline
$D=7$ & $\fft{SL(5,\sR)}{O(5)}$ & $\fft{SL(5,\sR)}{O(3,2)}$ \\ \hline
$D=6$ & $\fft{O(5,5)}{O(5)\times O(5)}$ & $\fft{O(5,5)}{O(5, C)}$
 \\ \hline 
$D=5$ & $\fft{E_{6(+6)}}{U\!Sp(8)}$ & $\fft{E_{6(+6)}}{U\!Sp(4,4)}$
 \\ \hline
$D=4$ & $\fft{E_{7(+7)}}{SU(8)}$  &  $\fft{E_{7(+7)}}{SU^*(8)}$ \\ \hline
$D=3$ & $\fft{E_{8(+8)}}{SO(16)}$ & $\fft{E_{8(+8)}}{SO^*(16)}$ \\ \hline
\end{tabular}
\end{center}

\bigskip

\centerline{Table 2: Cosets for maximal supergravities in
Minkowkian and Euclidean signatures}

\bigskip\bigskip

         As was discussed in \cite{lptdual,cjlp}, maximal
supergravities have global symmetries $E_{n(+n)}$ when all
dualisations that reduce the degrees of field strengths are performed.
If instead we dimensionally reduce $D=11$ supergravity to $D$
dimensions without performing any dualisations, then the resulting
theory will have a $GL(11-D,\R)\semi R^q$ global symmetry, where
$q=\ft16(11-D)(10-D)(9-D)$ \cite{lptdual,cjlp}.  From the point of
view of perturbative string theory, another natural possibility is to
dualise only R-R fields, since, unlike the NS-NS fields, they couple
to the world-sheet through their field strengths only.  If this is
done, the global symmetry becomes $O(10-D,10-D)\semi R^{8-D}$.  The
coset structures of these theories in Minkowskian and
Euclidean-signature spaces are given by

\bigskip\bigskip

\begin{center}
\begin{tabular}{|c|c|c|}\hline
 & Minkowskian & Euclidean \\ \hline
No-dual & $\fft{GL(11-D)\semi R^{\fft16(11-D)(10-D)(9-D)}}{O(11-D)}$ & 
$\fft{GL(11-D)\semi R^{\fft16(11-D)(10-D)(9-D)}}{O(10-D,1)}$ \\ \hline
RR-dual & $\fft{O(10-D,10-D)\semi R^{8-D}}{O(10-D)\times O(10-D)}$ &
$\fft{O(10-D,10-D)\semi R^{8-D}}{O(10-D, C)}$ \\ \hline
\end{tabular}
\end{center}

\bigskip

\centerline{Table 3: Cosets for non-dualised or RR-dualised
maximal supergravities}

\bigskip\bigskip

\section{Time reduction and type IIA/type IIB T-duality}

       So far we have discussed the dimensional reduction of
eleven-dimensional supergravity, in cases where one of the internal
directions is the time coordinate.  In this section we shall give an
analogous discussion for the type IIB theory, and re-examine the type
IIA/IIB T-duality when a time reduction is involved.

     The bosonic sector of type IIB supergravity comprises the metric,
a dilaton, a self-dual 5-form (with potential $B_4$), NS-NS and R-R
2-form potentials $(A_2^{\rm NS}, A_2^{\rm R})$, and one axion
$\chi$. The nine-dimensional Lagrangian that results from
dimensionally reducing this on a spatial $S^1$ can be found in
\cite{bho}.  In the case of a time reduction instead, it follows from
the discussion in section 2.1 that we need only modify the signs of
the kinetic terms for the 3-form potential coming from the reduction
of $B_4$, and all the vector potentials, since they are NC-fields.

\subsection{Type IIA/type IIB T-duality}

          First let us review the standard type IIA/type IIB T-duality
when the two theories are compactified on a spatial circle $S^1$.  The
relations between the gauge potentials of the two theories reduced to
$D=9$ are summarised in Table 4.

\bigskip\bigskip
\begin{center}
\begin{tabular}{|c|c|c|c|c|c|}\hline
    &\multicolumn{2}{|c|}{IIA} &
    &\multicolumn{2}{c|}{IIB} \\ \cline{2-6}
    & $D=10$ & $D=9$ &T-duality & $D=9$ & $D=10$ \\ \hline\hline
    & $A_\3$ & $A_\3$ & $\longleftrightarrow$ &
                   $\underline{A_\3}$ & $B_\4$ \\ \cline{3-6}
R-R & &  $\underline{A_{\2 2}}$& $\longleftrightarrow$
                           & $A_\2^{\rm R}$ & $A_\2^{\rm R}$
                                               \\ \cline{2-5}
fields& ${\cal A}_\1^{1}$ & ${\cal A}_\1^{1}$ &
                $\longleftrightarrow$ &
        $\underline{A_\1^{\rm R}}$ & \\ \cline{3-6}
   & & $\underline{{\cal A}^1_{\0 2}}$ & $\longleftrightarrow$
                            & $\chi$ &$\chi$
                                 \\ \hline\hline
NS-NS & $G_{\mu\nu}$ & $\underline{{\cal A}_\1^{2}}$
                        & $\longleftrightarrow$ &
        $\underline{A_\1^{\rm NS}}$ & $A_\2^{\rm NS}$ \\ \cline{2-5}
fields& $A_{\2 1}$ & $A_{\2 1}$ &
               $\longleftrightarrow$ & $A_\2^{\rm NS}$ &
                                       \\ \cline{3-6}
      & & $\underline{A_{\1 12}}$ & $\longleftrightarrow$ &
                              $\underline{{\cal A}_\1}$ & $G_{\mu\nu}$
                                       \\ \hline
\end{tabular}
\end{center}

\bigskip

\centerline{Table 4: Gauge potentials of type II theories in $D=10$
and $D=9$}
\bigskip\bigskip

       Note that the underlined fields are NC-fields (and therefore
have plus signs in front of their kinetic terms) if the reduction from
$D=10$ to $D=9$ is performed instead on the time coordinate.  The
relation between the dilatonic scalars of the two nine-dimensional
theories is given by
\be
\pmatrix{\phi_1 \cr \phi_2}_{\rm IIA} =
\pmatrix{\ft34 & -\ft{\sqrt7}{4} \cr
                                           -\ft{\sqrt7}{4} & -\ft34}
\pmatrix{\phi_1 \cr \phi_2}_{\rm IIB} \equiv M\, 
\pmatrix{\phi_1 \cr \phi_2}_{\rm IIB} 
\ .\label{dils}
\ee
Note that we have $M^{-1} = M$.   The dimensional reduction of 
the ten-dimensional string metric to $D=9$ is given by
\bea
ds_{\rm str}^2 &=& e^{\ft12\phi_1}\, ds_{10}^2 \nn\\
&=& e^{\ft12\phi_1}\, (e^{-\phi_2/(2\sqrt7)}\, ds_9^2 +
e^{\sqrt7\phi_2/2} \, (dz_2 + {\cal A})^2 ) \ ,
\eea
where $ds_{10}^2$ and $ds_9^2$ are the Einstein-frame metrics in
$D=10$ and $D=9$.  The radius of the compactifying circle, measured
using the ten-dimensional string metric, is therefore given by
$R=e^{\ft14 \phi_1 +\ft{\sqrt7}{4}\phi_2}$.  Note that the dilaton
vector $\{\ft14, \ft14 \sqrt 7\}$ of the radius is the eigenvector of
$M$ with eigenvalue $-1$.  It follows that the radii $R_{\rm IIA}$ and
$R_{\rm IIB}$ of the compactifying circles, measured using their
respective ten-dimensional string metrics, are related by $R_{\rm
IIA}=1/R_{\rm IIB}$.

        This picture of type IIA/IIB T-duality breaks down when the
theory is compactified instead on the time direction. In fact it is
non-perturbative states, such as D-branes, that can be held
responsible for this breakdown.  To see this, we first note that the
scalar coset manifold for the type IIB theory is $SL(2, \R)/O(2)$, and
this will remain as a factor in the complete scalar sector in $D=9$,
regardless of whether the compactification is on the time or a space
direction.  On the other hand, as we saw in section 3, the coset for
the Euclidean-signature $D=9$ type IIA theory has an $SL(2,\R)/O(1,1)$
factor.  (In each case, there is an additional scalar field that is
decoupled from the $SL(2,\R)$-invariant factor.) In other words, the
axionic scalars of the type IIA and type IIB theories in $D=9$
Euclidean-signature space have opposite signs for their kinetic terms.
In fact, it is easy to verify that all the R-R fields of the type IIA
theory in $D=9$ Euclidean-signature space (see Table 4) will have
opposite signs for their kinetic terms, in comparison to the kinetic
terms for the R-R fields of the Euclidean-signature $D=9$ type IIB
theory.  On the other hand in the NS-NS sector, the signs are in
agreement, since the vectors associated both with the Kaluza-Klein and
the winding modes (which are interchanged on passing between type IIA
and type IIB) acquire minus signs when the theories are compactified
on the time direction.\footnote{It is, of course, possible to make a
{\it complex} field redefinition in order to relate the two
nine-dimensional Euclidean-signature theories, by following the rules
given in Table 4, but with a factor of $i$ in the identification for
each R-R field.  Since this would therefore relate real solutions to
complex solutions both in $D=9$ and $D=10$, any ``T-duality'' would
have the undesirable consequence of requiring the existence of complex
solutions, even in the original Minkowskian-signature ten-dimensional
theories.}

          This sign discrepancy in the kinetic terms of the R-R fields
in the type IIA and type IIB theories, and hence the breakdown of the
T-duality, can also be understood from the point of view of D-brane
physics.  In the spatial $S^1$ compactification, a D$p$-brane in one
theory is dual to a D$(p+1)$-brane in the other theory, due to the
fact that in type IIA, D$p$-branes arise only for even $p$, while in
type IIB, they arise only for odd $p$.  In particular, this implies
that in going from $D=10$ to $D=9$, a D$(p+1)$-brane undergoes a
diagonal (double) dimensional reduction, where both the world-volume
and the spacetime dimension are reduced, while a D$p$-brane undergoes
a vertical reduction, in which both the transverse space of the brane
and the spacetime dimension are reduced.  If, on the other hand, we
compactify the theory on the time coordinate, then this means that
only diagonal dimensional reduction of $p$-brane solitons is
performed, since the time coordinate is always part of the
world-volume.  In other words, the time direction can participate only
in a diagonal reduction step, but not in a vertical reduction.  Since
the T-duality of D$p$-branes requires both double and vertical
reductions, it follows that the existence of these {\sl
non-perturbative} states leads to a breakdown of T-duality in the case
of a dimensional reduction in the time direction.

     As a cautionary note, it should be remarked that there do in fact
exist static $p$-brane solutions in which time is one of the
coordinates of the transverse space.  Such solutions can be vertically
dimensionally reduced on the time direction, giving rise to $p$-branes
in a Euclidean-signature space.  However, these solutions, and hence
also the original higher-dimensional solutions, will be complex.  The
reason for this can be seen most easily by looking in the reduced
theory; the field strength supporting the $p$-brane will have the
``wrong sign'' for its kinetic term.  Specifically, if the solution is
supported by an electric charge, then the reduced field strength has
the same degree as in the higher dimension, and so, by the results of
section 2.1, it will have a minus sign in its kinetic term.  On the
other hand, if the $p$-brane carries a magnetic charge, then the
reduced field strength will have a degree that is 1 less than in the
higher dimension, and hence its will have a plus sign in its kinetic
term.  In each case, the sign is the opposite of what is needed for a
real solution in a Euclidean-signature space.  For convenience, a
summary of the signs needed in order to have real solutions in
Minkowskian and Euclidean signature spaces is given in Table 5:

\bigskip\bigskip
 
\begin{center}
\begin{tabular}{|c|c|c|}\hline
   & Minkowskian & Euclidean \\ \hline 
Electric & $-F^2$  & $+F^2$ \\ \hline 
Magnetic & $-F^2$ & $-F^2 $ \\ \hline     
\end{tabular}
\end{center}

\bigskip

\centerline{Table 5: Signs of kinetic terms for real $p$-branes}

\bigskip\bigskip

   In the case of real solutions, an extremal $p$-brane satisfies the
usual relation $m=Q$ between the mass and the charge.  When the
solutions are instead complex, as a result of a wrong sign for the
kinetic term, the mass and charge are instead related by $m=iQ$.  It
would be natural in such cases to take the mass to be real, so that
the metric would be real, and therefore the charge would be imaginary.

     If one takes the point of view that all lower-dimensional
solutions are ultimately to be interpreted as solutions of the
original ten or eleven-dimensional theories, then it would be natural
to insist that all the solutions should be real.  On the other hand,
complex solutions might be regarded as being acceptable in
Euclidean-signature theories in their own right, since in fact any 
electric/magnetic dual pair of $p$-branes in a Euclidean-signature space  
will necessarily have one member that is complex, as can be seen from
Table 5.

    However, to return to our discussion of the two inequivalent
nine-dimensional Euclidean-signature theories, even if states with
imaginary charge were admitted in the spectrum, this would still not
imply a T-duality between the two theories, because it would require
the identification of states carrying real charges with states
carrying imaginary charges.   This would imply that the
ten-dimensional type IIA (or type IIB) theory would have complex
states.  Note that U-duality symmetries, which act transitively on the
charge lattice, will never map a solution with a real charge to a
solution with an imaginary charge.  In the case of the type IIA/IIB
T-duality, we should likewise expect that real solutions of the one
theory should map into real solutions of the other.  (Note again that
as we discussed in sections 2 and 3, the reality of a solution should
be judged by the reality of the charges and ${\cal M}$ (which are
physically observable), and does not necessarily require the reality
of the scalar fields themselves.)

      Purely within the NS-NS sector, this problem does not arise, even
for the duality between non-perturbative states such as 5-branes and
NUTs.  At the field-theory level, this is related to the fact that the
vector potentials coming from the 2-form potential and the metric both
have kinetic terms that undergo sign reversals when the reduction is
on the time direction. From the point of view of the $p$-brane
spectrum, T-duality requires that NS-NS strings or 5-branes (or waves
or NUTs) in the type IIA and type IIB theories either both undergo
vertical reduction, or both undergo diagonal reduction, in order to
match up in $D=9$, and so again no incompatibility arises.

       We may now examine how the NS-NS and R-R strings in the two
nine-dimensional Euclidean-signature theories transform under the
$SL(2,\R)$ global symmetry ($SL(2,\Z)$ at the quantum level).  This
symmetry acts transitively on the two-dimensional charge space of the
NS-NS and R-R strings.  However, it also has the effect in general of
transforming the scalar moduli.  There exists a (point-dependent)
denominator subgroup that leaves any chosen point in the modulus space
fixed.  Let us first consider the nine-dimensional Euclidean-signature
theory coming from the reduction of the type IIB theory on the time
direction, for which the scalar coset is $SL(2,\R)/O(2)$. Note that in
this case, the electric charges for NS-NS or R-R strings
will be imaginary, while their dual 4-branes will carry real magnetic
charges.  Without loss of generality, we may consider the string
solutions at the self-dual point $\tau_0=i$.  The $O(2)$ denominator
group is then of the form
\be
\pmatrix{\cos\theta &\sin\theta \cr -\sin\theta & \cos\theta}\ . \label{o2}
\ee
It has the effect of continuously rotating the NS-NS and R-R string (or
4-brane) charges $(q_1, q_2)$.  At the quantum level, the $O(2)$ group
reduces to its Weyl group $Z_2$ \cite{lpsweyl}, whose group elements
are given by (\ref{o2}) with $\theta=0$ and $\theta=\ft12\pi$.  The
Weyl group, which leaves the moduli invariant, has the effect of
making a discrete interchange between the NS-NS and R-R strings (or
4-branes).

         The picture is different for the nine-dimensional theory
coming from the reduction of the type IIA theory on the time
direction.  In this case, the NS-NS string (or the R-R 4-brane)
carries an imaginary charge, while the the R-R string (or the NS-NS
4-brane) carries a real charge.  It is no longer the O(2) group
(\ref{o2}) that leaves the self-dual point $\tau_0=j$ invariant, but
instead the $O(1,1)$ group
\be
\pmatrix{\cosh t &\sinh t \cr \sinh t & \cosh t}\ . \label{o11}
\ee
This group no longer acts as a rotation between NS-NS and R-R
charges.  In particular, a pure NS-NS solution can never be rotated to
a pure R-R solution, or {\it vice versa}.  At the quantum level, the
only surviving element of the $O(1,1)$ is just the identity,
corresponding to $t=0$ in (\ref{o11}).  Thus we see that the behaviour
of the NS-NS and R-R strings (or the magnetic dual 4-branes) under the
global $SL(2,\R)$ symmetry is very different in the two
nine-dimensional Eucludean-signature theories, coming from the
reduction of either type IIA or type IIB supergravity on the time
direction.

     Having shown that the type IIA and type IIB theories are not
equivalent when they are reduced on the time direction, it is worth
pointing out that they do become equivalent when they are further
reduced to $D=8$, by compactification on a spatial circle.  An easy
way to understand this follows from the fact that, as we showed in
section 2, the orders of time reduction and spatial reduction commute.
So the reduction first on time and then on a spatial $S^1$ is equivalent
to a reduction first on a spatial $S^1$ and then on time.  Since the
spatially-reduced theories are already equivalent in $D=9$, this
equivalence is then inherited by all the maximal supergravities in
$D\le 8$ Euclidean-signature spaces. Nevertheless, it is instructive to
look in detail at how the two inequivalent Euclidean-signature $D=9$
type IIA and type IIB theories become the equivalent when they are
further reduced on $S^1$ to $D=8$.

          To make the comparison, let us begin by considering the
dimensional reduction of type IIA and type IIB first spatially on
$z_2$, followed by a reduction on time $t=z_3$. (Note that we reserve
$z_1$ as the internal coordinate in the compactification of $D=11$
supergravity from $D=11$ to $D=10$.) Since the fields are already
identified in $D=9$, as given in Table 4, it follows that the
descendents of these fields are also identified in a one-to-one
fashion.  If instead we first compactify the type IIA and type IIB
theories on the time direction $z_2=t$, and then on the spatial
coordinate $z_3$, the identifications listed in Table 4 might seem
no longer to be applicable, since the R-R fields of the two theories
have opposite signs for their kinetic terms.  For example, the axion
$\chi$ of type IIB is a C-scalar whilst the field ${\cal A}^1_{\0 2}$
is an NC-scalar.  This seems to suggest that $\chi$ should be
identified with ${\cal A}^1_{\0 3}$, which is also a C-scalar.
However, the field strength for ${\cal A}^1_{\0 3}$ has a Kaluza-Klein
modification, namely ${\cal F}^1_{\1 3}= d{\cal A}^1_{\0 3} -{\cal
A}^2_{\0 3}\, d{\cal A}^1_{\0 2}$, whilst the field strengths for
$\chi$ and ${\cal A}^1_{\0 2}$ have no such modifications.  The
Kaluza-Klein modification implies that $\chi$ can only be identified
with ${\cal A}^1_{\0 2}$.  In order to resolve the sign discrepancy of
the two kinetic terms, we need to perform a field redefinition of the
type given in (\ref{phitrans}), which has the effect of reversing the
sign of the kinetic term for ${\cal A}^1_{\0 2}$.  (Note that the sign
of the kinetic term for $\chi$ cannot be changed, as discussed in
section 3, since it is simply the one inherited from the $\chi$
kinetic term in $D=10$ type IIB.)  This transformation on $\vec\phi$
effectively interchanges the order of the reduction on the two
coordinates, so that it becomes first a spatial reduction, followed by
the time reduction.

    We should therefore identify the fields $\chi$ and ${\cal A}^1_{\0
2}$ in $D=8$, even though the former is a C-scalar while the latter is
an NC-scalar in the reduction from $D=10$ first on time and then on
space.  This immediately raises an apparent paradox, in which one
might think that we could just as well have done the same thing
already in nine dimensions, since ${\cal A}^1_{\0 2}$ is a field that
already exists in $D=9$.  However, there is a subtle difference
because in $D=8$, owing to the existence of the extra dilaton, there
is a freedom to perform a field redefinition (\ref{phitrans}) whose
effect is to reverse the sign of the kinetic term of ${\cal A}^1_{\0
2}$.  This is not possible in $D=9$, because although any field
redefinition of the form (\ref{phitrans}) has the effect of altering
which fields are compact and which non-compact, it is always in a way
that preserves the total number of NC field strengths and the total
number of C field strengths of each degree.\footnote{In the case of
field strengths of degree $n$ in $D=2n$ dimensions, the counting of
the C and NC fields should include their Hodge duals, since the field
strengths and their duals are both included in a single irreducible
representation of the global symmetry group.  For example, $F_\4$ and
its Hodge dual in Euclidean-signature $D=8$ supergravity form a
doublet under $SL(2,\R)$, one component of which is C, while the other
is NC.  For this reason although a field redefinition of the type
given in (\ref{phitrans}), but for the case of the type IIB reduction
to $D=8$, reverses the sign of the kinetic term for $F_\4$, this does
not contradict the rule that the total numbers of C and NC fields are
preserved.}  In $D\le 8$ there exists more than one axion, and hence
(\ref{phitrans}) can be used to redefine which axions are NC and which
are C, while keeping the total numbers of each fixed.  In $D=9$
however, ${\cal A}^1_{\0 2}$ is the only axion, and it is non-compact
in the Euclidean-signature space.  Thus a field redefinition of the
form (\ref{phitrans}), (which preserves the reality of the physical
quantities such as the matrix ${\cal M}$) cannot alter this
non-compactness, and hence ${\cal A}^1_{\0 2}$ cannot be identified
with the compact scalar $\chi$ coming from the type IIB theory.  So it
is only by descending one step further, with a spatial
compactification to $D=8$, that the identification of the type IIA and
type IIB fields can be effected.

\section{Instantons in $SL(2,\R)/O(1,1)$ Lagrangians}

         Scalar cosets coupled to gravity can support real electric
instanton solutions in Euclidean-signature spaces.  In this section,
we study instantons in the simplest non-trivial scalar coset, namely
$SL(2,\R)/O(1,1)$.  We study the orbits of the extremal instanton
solution, and show that an $SL(2,\R)$ transformation does not alter
its essential form, except for a constant shift and rescaling of the
harmonic function that characterises the solution.  Since such an
instanton in $SL(2,\R)/O(1,1)$ can be obtained from dimensional
reduction of any $p$-brane on its world-volume, it follows that this
duality symmetry relates any $p$-brane solution to its near-horizon
limit.  We also construct non-extremal instanton solutions and
conclude that their existence requires at least a $GL(2,\R)$ invariant
scalar-manifold, extending the results described above for extremal
instantons.

\subsection{Orbits of extremal instantons}

       The Lagrangian (\ref{sl2rlag1}) for the scalar coset
$SL(2,\R)/O(1,1)$, together with the Einstein-Hilbert term, admits an
extremal instanton solution in $D$-dimensional Euclidean space
\cite{dkl,stainless,ggp}:
\bea
ds^2 &=& dr^2 + r^2 d\Omega^2\ ,\nn\\
e^{\phi} &=& H\ ,\qquad \chi = H^{-1}\ ,\label{sol1}
\eea
where $H$ is an harmonic function on the Euclidean space.  For the
purpose of our discussion, we may consider an isotropic solution,
namely $H=1 + Q/r^{D-2}$.  The asymptotic values of the scalars
$\tau_0=\chi_0 + j e^{-\phi_0}$ for this solution are given by
$\tau_0=1 + j$.  (The solution at the self-dual point $\tau_0=j$ can
be obtained by shifting the axion $\chi$ to $\chi =H^{-1} -1$, and
indeed solutions at any other modulus point can be obtained by making
constant shifts of the dilaton $\phi$ and the axion $\chi$, using the
Borel subgroup of $SL(2,\R)$ transformations.)  The $\chi$ and $\phi$
fields in the solution (\ref{sol1}) can be combined to give $\tau=
H^{-1} (1+j)$.  Applying the $SL(2,\R)$ transformation (\ref{sl2rtr}),
we find
\be
\tau'= \fft{a d + bc + bd H + j}{2c d + d^2 H}\ .
\ee
Thus we obtain the new solution
\be
e^{\phi} =H'\equiv 2cd + d^2 H\ ,\qquad
\chi = {H'}^{-1} + \fft{b}{d}\ .\label{newh}
\ee
We see that the structure of the solution is unchanged, except for a
constant shift and rescaling of the harmonic function $H$.\footnote{
This result was also obtained for D-instantons in \cite{bb}.}
Later, in section 5.3, we shall show that this ability to shift the
constant term in the harmonic function can be used in order to relate
any extremal $p$-brane to its near-horizon limit, using the relevant
$SL(2,\R)$ subgroup of the duality group, in the dimension where the
$p$-brane has been reduced to an instanton.  We shall show that it can
also be done for multi-charge $p$-branes, and intersections, and also
that similar transformations can be made in the case of non-extremal
$p$-branes.

       The trivial modification of the harmonic function of the
instanton solution under the full $SL(2,\R)$ transformation suggests
that the instanton is a singlet.  In fact, we may now show that the
instanton is a singlet under the discretised $SL(2,\Z)$ symmetry of
the quantum theory.  We may illustrate this by examining the orbits of
the charges of the instanton under the $SL(2,\R)$ and $SL(2,\Z)$
transformations.  Instantons carry electric charges, which can be
defined to be the integrals of the duals of the Noether currents for
the global symmetries of the scalar coset.  There are three Noether
currents for the Lagrangian (\ref{sl2rlag1}), with its
$SL(2,\R)/O(1,1)$ coset:
\be
J_0=-d\phi - e^{2\phi}\chi d\chi\ ,\quad
J_{+} = e^{2\phi} d\chi\ ,\quad
J_{-} = -d\chi - 2\chi d\phi - e^{2\phi} \chi^2 d\chi
\ .\label{sl2rcurrents}
\ee
(See appendix A for a derivation of Noether currents for
arbitrary scalar coset manifolds.)
$J_0$ and $J_{+}$ can be called Borel currents, since they are
associated with the shift symmetries of the two scalars, which are
generated by the Borel subgroup of $SL(2,\R)$.  The $J_{-}$ current,
which can be expressed as a linear combination of $J_0$ and $J_{+}$
with scalar-dependent coefficients, is associated with transformations
generated by the negative root.   These three Noether currents 
transform linearly under the  adjoint representation of $SL(2,\R)$:
\be
{\cal J} \longrightarrow {\cal J}' = \Lambda^{-1}\,  {\cal J}\,  \Lambda
\ ,\label{sl2rctr}
\ee 
where
\be
{\cal J} = \pmatrix{J_0 & J_{-} \cr J_{+} & - J_0}\ ,\label{sl2noether}
\ee
and $\Lambda$ is a constant $SL(2,\R)$ matrix.  The charges of the
instanton then can be defined as
\be
{\cal Q} =\int *{\cal J} =\pmatrix{ Q_0, & Q_{-}\cr Q_{+} & -Q_0}
\ ,
\ee
which therefore transform in the same way as the Noether currents
${\cal J}$.

      The standard global symmetry group $SL(2,\R)$ transforms not
only the charges, but also the scalar moduli, \ie the asymptotic
values of the scalar fields at infinity.  For any point in the modulus
space, there exists a (modulus-dependent) $O(1,1)$ stability subgroup
that leaves the modulus fixed.  We shall examine how the charges
transform under this denominator subgroup.  Without loss of
generality, we may consider the instanton solution at the self-dual
point $\tau_0=j$, {\it i.e.}\ $e^{\phi} = H$ and $\chi = H^{-1} -1$.
Substituting this into the expression (\ref{sl2noether}) for the
Noether currents we find that
\be
{\cal J} = \pmatrix{-dH & dH \cr
                    -dH & dH}\ ,\label{sl2rinstnot}
\ee
and hence the Noether charges are 
\be
{\cal Q} = \pmatrix{-Q&  Q \cr -Q & Q}\ ,\label{sdcharge}
\ee
where $Q=\int *dH$.  Note that the 
$SL(2,\R)$-invariant quadratic quantity ${\rm Det}({\cal
Q})$ vanishes for the instanton solutions.  The $O(1,1)$
transformation at the self-dual point $\tau_0=j$ is given by
\be
\Lambda_{O(1,1)} =\pmatrix{\cosh t &\sinh t \cr
                           \sinh t &\cosh t }\ ,\label{sl2ro11}
\ee
and it has the effect of simply rescaling charges:
\be
{\cal Q} \longrightarrow {\cal Q}' = e^{-2t} \, {\cal Q}\ .
\ee
Thus we see the classical symmetry group $O(1,1)$ does not rotate the
charges in the charge lattice; rather it merely rescales the charges.
In fact it has the same effect on the charge lattice as does the
``trombone'' symmetry \cite{trombone}, under which the metric is 
rescaled: $g_{\mu\nu} \rightarrow \lambda^2 g_{\mu\nu}$.

        At the quantum level the $O(1,1)$ degenerates to the identity
group, and hence the charge cannot be changed.  This implies that the
instanton solution is a singlet under the $SL(2,\Z)$ symmetry.  At
this point, it is instructive to compare the instanton solution with
the usual $p$-branes supported by higher-degree field strengths.  For
such $p$-branes, the charges carried by the participating field
strengths are independent parameters. In other words, for any given
choice of scalar moduli, there exist solutions whose charges fill out
a charge lattice.  In this case, it is necessary to find a
spectrum-generating symmetry that maps between the solutions whose
charges lie at different points in the charge lattice, while holding
the moduli fixed.  It was shown in \cite{trombone} that this can be
done by means of a non-linearly realised duality symmetry whose
action is quite distinct from that of the standard linear action of
the global supergravity symmetry on the higher-degree fields, and in
particular it makes essential use of the trombone rescaling symmetry
too.  In our present case where we are considering instead instanton
solutions, the charges are not independent parameters; instead they
are related to the modulus parameters (up to trombone rescalings).
Thus for any given choice of scalar moduli, there is only one charge
configuration.  For example, at the self-dual point the charge of any
instanton solution has the same form as the one given in
(\ref{sdcharge}).  Any $SL(2,\R)$ transformation that had the effect
of rotating the charges would necessarily also change the values of
the scalar moduli.  This shows that the instanton solutions in the
$SL(2,\R)/O(1,1)$ theory are singlets under the spectrum-generating
symmetries.

\subsection{Non-extremal instantons}

         In the previous subsection, we discussed extremal instanton
solutions in the $SL(2,\R)/O(1,1)$ system in a Euclidean-signature
space.  Naively one would expect that as in the cases of general
$p$-branes, these solutions should be straightforwardly generalisable
to non-extremal instanton solutions.  In this subsection we show that
in fact the $SL(2,\R)/O(1,1)$ coset cannot support a non-extremal
instanton that is {\it isotropic} in the transverse space.  In fact an
isotropic non-extremal instanton requires the use of an additional
scalar field, meaning that we require a $GL(2,\R)/O(1,1)$ scalar
manifold in order to be able to describe it.

      A non-extremal instanton in $D$ dimensions can be obtained from
the dimensional reduction on the time direction of a non-extremal
static black hole in $D+1$ dimensions.  But first let us discuss the
dimensional-reduction properties of more general non-extremal
$p$-branes.  For simplicity, we consider first the single-charge
non-extremal $p$-brane solution in maximal supergravity in $(D+1)$
dimensions that involves a single $n$-form field strength with dilaton
vector $\hat {\vec c}$.  The relevant part of the Lagrangian describing 
the solution is then given by
\be
{\hat e}^{-1} {\cal L}_{D+1} = \hat R -\ft12 (\del \hat \phi)^2 -
\fft1{2n!} e^{\hat a\hat \phi} \hat F_n^2 \ ,\label{nformlag}
\ee
with $\hat \phi = \hat{\vec c} \cdot \hat {\vec\phi}$ and $\hat
a=|\hat{\vec c}|$.  
Note that for all such single-charge $p$-branes, we have $\hat a^2 = 4 -
2(n-1)(D-n)/(D-1)$.    The Lagrangian allows electric non-extremal 
$(n-2)$-brane solutions, given by \cite{dlpblack,ct}
\bea
&&ds_{D+1}^2 = -e^{2A'} dt^2 + e^{2A} dx^i dx^i +
                e^{2B} (e^{-2f} dr^2 + r^2d\Omega^2)\ ,\nn\\
&&e^{2A'} = e^{2A +2f}\ ,\qquad
e^{2A} = H^{-\ft{\td d}{D-1}}\ ,\qquad
e^{2B} = H^{\ft{d}{D-1}}\ ,\label{blackp}\\ 
&&F_n = \coth\mu\, dH^{-1} \wedge d^d x\ , \qquad \hat\phi = \ft12
\hat a \log H \ ,\nn
\eea
where
\be
H=1 + \fft{k\sinh^2\mu}{r^{\td d}} \ ,\qquad 
e^{2f} = 1 - \fft{k}{r^{\td d}}\ ,\label{hf}
\ee
and $d=n-1$, $\td d = D-n$.  Here $k$ and $\mu$ are constants,
parameterising the charge, $Q=\ft12 k\, \sinh 2\mu$ and mass per unit
$p$-volume $m= k\, (\td d\, \sinh^2\mu + \td d+1)$ (the extremal limit
is achieved by sending $\mu$ to infinity and $k$ to zero, holding $k\,
e^{2\mu}$ constant).  Note that we can write $\hat F_n$ in terms of
its potential $\hat A_{n-1}$, with
\be
\hat A_{n-1}= \cosh\mu\,\, H^{-1} \wedge d^dx  \ .
\ee
The solution (\ref{blackp}) has an important property, namely
\be
A\, d  +B\,  \td d  =0\ . \label{abrelation}
\ee
Here we shall continue to refer the function $H$ in (\ref{hf}) as an
`harmonic' function.  The Lagrangian (\ref{nformlag}) also admits a
magnetic $(D-n-2)$-brane in $(D+1)$ dimensions which we shall not
consider further, since the discussion is analogous to that for the
electric solution.  Let us now consider the diagonal dimensional
reduction on a world-volume coordinate of the non-extremal $p$-brane
in $(D+1)$ dimensions to a $(p-1)$-brane in $D$
dimensions.\footnote{{\it Vertical} dimensional reduction of
non-extremal $p$-branes requires the construction of an infinite
number of non-extremal $p$-brane in $(D+1)$ dimensions, periodically
arrayed along the internal coordinate $z$ that is to be compactified.
The symmetry associated with the periodicity implies the equilibrium
of the configuration, and the compactification of $z$ implies the
stability.  See \cite{lpx}.} If we dimensionally reduce the
electrically-charged solution on one of the world-volume spatial
coordinates $x^i$, the supporting $D$-dimensional field strength will
become an $(n-1)$-form, and so the relevant $D$-dimensional Lagrangian
will be
\be
e^{-1} {\cal L}_D = R -\ft12(\del\phi)^2 -\ft12 (\del\varphi)^2
-\ft1{2(n-1)!} e^{-2(D-n)\a\varphi + \hat a \hat\phi} F_{n-1}^2
\label{dnm1formlag}
\ee
where $\a = 1/(2(D-1)(D-2))^{1/2}$.  It was observed in \cite{lpsvert}
that when the condition $A\, d + B\, \td d =0$ is satisfied by the
$p$-brane solution in $D+1$ dimension, the linear combination of
dilatons $a\phi = -2(D-n)\a\varphi + \hat a \hat \phi$ (with $a^2=\hat
a^2 + 4(D-n)^2\a^2$) that couples to $F_{\sst(n-1)}$ in $D$ dimensions
is non-vanishing, whilst the orthogonal linear combination vanishes.
This means that a single-charge non-extremal $p$-brane in $D+1$
dimensions reduces to a standard single-charge single-scalar
non-extremal $(p-1)$-brane in $D$ dimensions.  (The discussion for the
diagonal reduction of the magnetic solution is analogous, in which
case the field strength $F_\n$ in the lower-dimensional theory will be
the relevant one that supports the $(p-1)$-brane, and its dilaton
coupling will be non-vanishing whilst the orthogonal dilaton
combination will vanish, leading again to a standard single-charge
single-scalar solution.)  If we instead reduce the $p$-brane solution
on the time direction, then in the {\em extremal} case we have $e^{2f}
=1$, and hence the conclusion is the same as for reduction on a spatial
world-volume direction.  However, if we start the timelike
dimensional reduction from a {\it non-extremal} $p$-brane solution we
have, from (\ref{blackp}), $A'\, d + B\, \td d \ne 0$, which implies that
the other combination of the two dilatons, orthogonal to the combination
that couples to the field strength, will also be non-vanishing in $D$
dimensions.  Thus non-extremal $p$-brane solutions in
Euclidean-signature spaces are supported by a set of fields that
includes an additional scalar, which does not couple to the field
strength that carries the charge.

       In our present case, we are particularly interested in the
non-extremal instanton solutions that can be obtained from the
dimensional reduction of non-extremal black holes, which arise as
solutions for the $(D+1)$-dimensional Lagrangian (\ref{nformlag}) with
$n=2$; for example, D0-branes in $D=10$ type IIA theory.  The
non-extremal black hole solution is given by (\ref{blackp}) with
$d=1$, $\td d =D-2$.  From the Kaluza-Klein ansatz
\be
ds_{D+1}^2 = e^{-2\a\varphi} ds_{D}^2 - e^{2\a(D-2)\varphi} dt^2
\ ,\qquad \a = \fft{1}{\sqrt{2(D-1)(D-2)}}\ ,
\ee 
(the Kaluza-Klein vector is zero in this case) and from
(\ref{blackp}), we see that
\be
e^{2(D-2)\a\varphi} = e^{2f}\, H^{-\ft{D-3}{D-2}}
\ee
for the reduction of the black hole to an instanton in $D$ dimensions.
This instanton will be of a solution of the equations following from
the reduced Lagrangian
\bea
e^{-1} {\cal L}_D &=& R - \ft12(\del\hat \phi)^2 -\ft12(\del \varphi)^2
+\ft12 e^{-2(D-2)\a\varphi + \hat a\hat\phi} (\del\chi)^2\nn\\
&=& R -\ft12 (\del \phi_2)^2 -\ft12 (\del \phi_1)^2 +
e^{2\phi_1} (\del\chi)^2\ ,\label{gl2rlag}
\eea
where
\be
\phi_1 = \ft12 \hat a\hat \phi -(D-2)\a\, \varphi\ ,\qquad
\phi_2 = (D-2)\a\, \hat\phi + \ft12\hat a \varphi\ .
\ee
The axion $\chi$ results from the dimensional reduction of
$A_1$, according to $A_1 \rightarrow \chi dt$.  Thus in terms of these
new variables the non-extremal instanton solution in $D$-dimensions for
the Lagrangian (\ref{gl2rlag}) is given by
\bea
&&ds^2 = e^{\ft{2f}{D-2}}(e^{-2f} dr^2 + r^2 d\Omega^2)\ ,\nn\\
&&\phi_1 = -f + \log H\ ,\qquad \chi= H^{-1} \coth\mu\ ,\qquad
\phi_2 = f \sqrt{\ft{D}{D-2}}\ .
\eea

     We see that the existence of a non-extremal instanton requires a
Lagrangian containing at least two dilatons, and thus at least a
$GL(2,\R)\sim \R\times SL(2,\R)$ invariant scalar manifold, although
the $\R$ factor decouples in the extremal limit.  This phenomenon may
be of significance in the understanding of an F-theory interpretation
\cite{vafa} of the type IIB theory.  The ten-dimensional type IIB
theory has only an $SL(2,\R)$-invariant scalar Lagrangian.  The
extremal instanton solution of the Euclideanised theory was
constructed in \cite{ggp}.  Its twelve-dimensional interpretation as a
pp-wave was put forward in \cite{tseylinwave}.  In this case, the
scalar field associated with the volume of the two-torus in the
compactification of F-theory to type IIB was considered to be
non-dynamical \cite{vafa}\footnote{It was shown in \cite{kklp} that it
cannot simply be taken to be non-dynamical once the higher-degree
fields of the theory are included.}, and indeed it decouples in the
extremal instanton solution.  However, this scalar associated with
volume of the 2-torus would have to be non-zero in a non-extremal
instanton.

        In \cite{tseylinwave}, a non-extremal instanton solution
within the $SL(2,\R)$ system was constructed for the type IIB theory,
using the T-duality that maps D0-branes in type IIA to instantons in
type IIB.  Putting aside for now the previously-noted obstacles to the
implementation of type IIA/IIB T-duality on the time direction, we may
note also that the non-extremal instanton constructed in
\cite{tseylinwave} is not isotropic in the ten-dimensional
Euclidean-signature space.  In other words, the solution has a $U(1)$
isometry along the Euclideanised time axis $y_0$, and hence the
harmonic function $H$ is given by $1 + Q (y_1^2+\cdots y_9^2)^{-7/2}$
rather than $1 + Q (y_0^2+\cdots y_9^2)^{-4}$. As we showed above, the
{\it isotropic} non-extremal instanton does not exist in a system with
only an $SL(2,\R)/O(1,1)$ scalar manifold; it requires an additional
independent scalar in order to support the solution.  If the type
IIA/IIB T-duality also implied a relationship between non-BPS states
such as non-extremal $p$-branes, then the consequent existence of a
non-extremal isotropic instanton in type IIB would give supporting
evidence for the existence of F-theory, since the emergence of the
necessary extra scalar could easily be understood from a
twelve-dimensional point of view, in parallel to the relation of
nine-dimensional non-extremal instantons to non-extremal pp-waves in
$D=11$.

        Since the non-extremal instanton solution is described by a
$GL(2,\R)/O(1,1)$ scalar Lagrangian, we may examine how it transforms
under the $GL(2,\R)\sim \R \times SL(2,\R)$ global symmetry.  The $\R$
factor of transformation is straightforward, implying simply a
constant shift of the scalar $\phi_2$.  On the other hand, the global
$SL(2,R)$ transformations act on the $(\phi_1, \chi)$ system in the
standard way, while leaving $\phi_2$ invariant.  Defining the
double-number valued field $\tau = \chi + j\, e^{-\phi_1}$, then under
fractional linear transformations $\tau\rightarrow (a\, \tau + b)/(c\,
\tau + d)$, we find that the fields in the instanton solution
transform to
\be
\phi_1' = -f +  \log H' \ ,\qquad \chi' =
\Big(\coth\mu + \fft{c}{d}\, {\rm cosech}^2\mu \Big)\, H'{}^{-1} +
\fft{b}{d}\ ,
\ee
where the transformed harmonic function $H'$ is given by
\be
H' = d^2\, H + c^2\, {\rm cosech}^2\mu + 2c\, d\, \coth\mu\ ,
\label{nonextremaltr}
\ee
while the field $\phi_2$ remains unchanged.  Thus we see that as in
the extremal case, the form of the solution is the same as before
except that the harmonic function $H$ is rescaled and shifted by a
constant.  This extends the previous result for extremal
D-instantons \cite{bb} to include arbitrary non-extremal instantons.

       So far we have considered just single-charge instanton
solutions, obtained by dimensionally reducing single-charge
non-extremal $p$-branes on the entire set of $d=p+1$ world-volume
directions.  More generally, if the Lagrangian in $D$ dimensions
contains a number of $n$-form field strengths, for which a subset of
$F_\n^\a$ ($\a=1,\ldots,N$) have dilaton vectors satisfying the 
dot-product relations \cite{lpmulti}
\be
\hat {\vec c_\a}\cdot\hat{\vec c_\beta} = 4 \delta_{\a\beta} -
\fft{2d \, \td d}{D-2}\ ,
\ee
then there exist $N$-charge non-extremal $p$-branes \cite{dlpblack}.
These solutions can be diagonally reduced to give $N$-charge non-extremal
instantons, which are solutions of the equations of motion following
from the Lagrangian
\be
e^{-1}{\cal L} = R - \ft12({\del\varphi})^2 -\ft12 (\del\vec\phi)^2 +
\ft12 \sum_{\a=1}^N e^{\vec c_\a \cdot \vec \phi} (\del\chi_\a)^2
\ ,\label{nchargelag}
\ee
where we have $\vec c_\a \cdot \vec c_\beta =4\delta_{\a\beta}$.  The
proof of the absence of Kaluza-Klein modifications and ${\cal
L}_{\sst{FFA}}$ terms can be found in \cite{classp}. Thus
we see that the dilatons $\varphi$ and $\varphi_\a \equiv\vec
c_\a\cdot \vec \phi$ are completely decoupled from each other, and the
pairs $(\varphi_\a, \chi_\a)$ form a total of $N$ independent
$SL(2,\R)/O(1,1)$ cosets.  The non-extremal instanton solution in $D$
dimensions is then given by
\bea
ds^2 &=& e^{\fft{2f}{D-2}}\, (e^{-2f}\, dr^2 + r^2 \, d\Omega^2)\ ,\nn\\
\varphi_\a &=& -f +  \log H_\a\ ,\qquad \chi_\a = H_\a^{-1}\, \coth\mu_\a\ ,
\qquad \varphi = f\, \sqrt{\ft{D}{D-2}} \ ,
\eea
with $H_\a = 1 + (k\sinh\mu_a)\, r^{-(D-2)}$.  Acting with the
independent $SL(2,\R)$ transformations on the $N$ cosets, we are able
to make independent transformations of the form (\ref{nonextremaltr})
on each of the harmonic functions $H_\a$.  Note that in the extremal
limit $k\rightarrow 0$ we have $f \rightarrow 0$, and hence the extra
scalar $\varphi$ decouples from the system.  In this case, the
harmonic functions $H_\a$ can each be independently shifted and
scaled, as in (\ref{newh}), so that they become
\be
H_\a' = 2 c_\a\, d_\a\ + d_\a^2\, H_\a\ ,
\ee
under the $SL(2,\R)$ transformations
\be
\Lambda_\a =\pmatrix{ a_\a& b_\a\cr
                      c_\a &d_\a}\ .
\ee 

\subsection{Instanton transformation and $p$-brane asymptotic geometry}

       As we have seen in the previous two subsections, a generic
$SL(2,\R)$ transformation does not change the essential structure of
the instanton solution, but it does have the effect of modifying the
harmonic function $H$ by a constant shift and a constant rescaling.
This is true both for the extremal and the non-extremal instantons.
For particular choices of the $SL(2,\R)$ transformation parameters,
however, this can have the effect of completely altering the
asymptotic behaviour of the harmonic function in the limit where
$r\rightarrow \infty$.  Thus if the $SL(2,\R)$ parameters are chosen
so that $c=-\ft12 d$, the harmonic function $H'$ defined in
(\ref{newh}) becomes
\be
H' =  \fft{Q\, d^2}{r^{D-2}}\ .
\ee
This duality transformation has therefore had the effect of expanding
the near-horizon structure of the instanton, where the $Q\, r^{2-D}$
term in $H$ dominates over the constant term, to the entire range of
$r$ values.

     Let us now consider this phenomenon in more general situations,
in which we begin by considering a $p$-brane solution in a higher
dimension. To begin with, we shall take the example of a single-charge
extremal $p$-brane.  This solution can be diagonally dimensionally
reduced through a succession of steps, until we arrive at a
single-charge instanton, after having compactified the entire
$(p+1)$-dimensional world-volume of the $p$-brane.  This will be a
solution that involves only a subsector of the lower-dimensional
supergravity fields, namely the metric, a certain single combination
of the dilatons, and the axion whose field strength carries the charge
supporting the instanton.  This dilaton/axion pair will be described
by a standard $SL(2,\R)/O(1,1)$ coset. Since the reduction was
performed on the world-volume, the harmonic function $H=1 + Q r^{-\td
d}$ of the higher-dimensional $p$-brane solution, with $\td d=D-3-p$,
is exactly the same as the harmonic function of the instanton solution
in the lower dimension.

    We now perform the $SL(2,\R)$ transformation on the instanton, as
described above, and obtain the new harmonic function $H'= Q' r^{-\td
d}$.  Having done so, we retrace the previous steps and diagonally
oxidise this transformed instanton solution back to the original
higher dimension.  Thus we arrive at an extremal single-charge
$p$-brane solution that differs from the original one only in having
the original harmonic function $H$ replaced by $H'$.  The asymptotic
structure of the new $p$-brane solution with $H'$ is therefore
altered, and now takes the same form as the near-horizon limit of the
original solution, in the regime where the constant term in $H$ was
negligible in comparison to $Q r^{-\td d}$.  Cases of particular
interest arise when the dilaton in the original $D$-dimensional
$p$-brane solution is finite on the horizon, since the near-horizon
structure then approaches AdS$_{(p+2)}\times S^{D-p-2}$ and the
supersymmetry is enhanced \cite{gt,dgt,ght,berg,kk}.  In such cases,
the $SL(2,\R)$-transformed solution has the global structure of
AdS$_{(p+2)}\times S^{D-p-2}$ everywhere.  (See also \cite{cowt,bps2}
for non-standard intersections \cite{khuri,bbj,gkt} that give rise to
AdS structures.)  

      At first sight, there might seem to be a paradox regarding this
enhancement of supersymmetry, since the $SL(2,\R)$ symmetry of the
supergravity theory describing the instanton is a subgroup of the
U-duality group.  U-duality commutes with supersymmetry, and hence one
might expect that the $SL(2,\R)$ transformation of the solution should
leave its supersymmetry unchanged.  The paradox is resolved by the
observation that the AdS space can be viewed as a special case of a
$(D-2)$-brane, or domain wall solution, in horospherical coordinates.
Half the Killing spinors in this description of AdS depend on the
world-volume coordinates \cite{lpt}, and so this half of the
supersymmetry in AdS is lost when the solution is reduced to an
instanton (since in the reduction process it is assumed that none of
the fields depends on the world-volume coordinates).  This explains
why the fraction of preserved supersymmetry in a $p$-brane with an
AdS$\times$Sphere near-horizon structure is always doubled at the
horizon.

        This discussion can be extended also to the discretised $SL(2,\Z)$
U-duality group of the quantum theory.  This shows that the
near-horizon geometry captures the essence of any $p$-brane, since it
is dual to the $p$-brane itself.

   A generalised discussion can be given for any harmonic intersection
of $p$-branes, waves and NUTs.  The solution for the intersection of
$N$ basic objects involves $N$ independent harmonic functions $H_\a$.
These solutions can be dimensionally reduced to $N$-charge $p$-branes,
which can then be further reduced to $N$-charge instantons.  The
nature of the harmonic intersection implies that the participating
fields that support this $N$-charge instanton can be described by a
Lagrangian of the form (\ref{nchargelag}), where $\varphi$ decouples
in the extremal limit.  In particular the dilaton vectors satisfy
$\vec c_\a \cdot \vec c_\beta = 4\delta_{\a\beta}$.  This implies that
the dilatons $\varphi_\a=\vec c_\a \cdot \vec \phi$ are decoupled from
each other, and the system is described by $N$ independent
$SL(2,\R)/O(1,1)$ cosets.  Each $SL(2,\R)$ factor in the total global
symmetry group can therefore be used to transform the associated
harmonic function.  This means that for appropriate choices of the
various $SL(2,\R)$ parameters, the constant terms in all $N$ harmonic
functions can be independently adjusted, and in particular, caused to
vanish.  If the the harmonic functions are chosen to have charges that
are equal and coincident, then the solution becomes a bound-state
$p$-brane with $\Delta =4/N$.  Thus we have seen that using
dimensional reduction on the time direction, and $SL(2,\R)$
transformations of the instantons, we can alter the asymptotic
geometry for all $p$-branes, intersections and bound states.

          The above discussion applies equally well to non-extremal
$p$-branes, whose dimensional reductions on their world-volumes give rise
to non-extremal instantons.  The $SL(2,\R)$ transformation leave the
non-extremal factor $e^{2f}$ invariant, but can rescale and shift the
`harmonic' function $H$ by constants, as given in
(\ref{nonextremaltr}).

    The use of duality transformations to alter the asymptotic
structure of $p$-branes was first proposed in \cite{hyun}, and
developed in \cite{bps}, by utilising a combination of general
coordinate transformations, the S-duality symmetry in type IIB, and
the T-duality transformation between the type IIA and type IIB
theories.  In general, the prescription in \cite{bps} is to start from
a $p$-brane in $D$ dimensions, map it to $D=10$ by oxidation or
reduction, and then by then by means of a sequence of T-duality
transformations (together with an S-duality symmetry transformation if
the starting point was an R-R $p$-brane), map it to a ten-dimensional
wave solution.  Next, a linear coordinate transformation mixing the
time and the longitudinal wave directions is performed, which has the
effect of shifting and rescaling the harmonic function by constants.
Finally, the previous sequence of duality and oxidation or reduction
steps is retraced, eventually giving back a $p$-brane with a shifted
and rescaled harmonic function.  A succession of such processes can be
performed in order to make independent shifts and rescalings of all
the harmonic functions in a multi-charge $p$-brane or an intersection.
In particular, this procedure was used to relate non-dilatonic black
holes in $D=5$ and $D=4$ to their near-horizon $AdS_3\times {\rm
Sphere} \times {\rm Torus}$ structures.  This provides a conformal
field theoretic understanding of the entropy of non-dilatonic black
holes \cite{ss}.

   Recently, a different procedure for shifting and scaling the
harmonic functions in certain $p$-brane solutions was proposed
\cite{bb}.  Effectively, the idea is to follow similar steps to those
described above, except that the goal is to map the $p$-brane into the
instanton of the type IIB theory, rather than to map it to a wave in
$D=10$.  Now, one uses the $SL(2,\R)$ symmetry of the type IIB theory
to shift and scale the harmonic function, in a similar manner to the
procedure we described in section 5.1.  Again, by retracing the
dualisation and oxidation or reduction steps, one arrives at a
$p$-brane with whose harmonic function is shifted and scaled relative
to the starting point.  There is an interesting question that can be
raised about this procedure, since in order to describe instantons in
the type IIB theory in $D=10$ it is necessary to Euclideanise the
theory.  As we discussed in section 2, the issue arises of how the
Majorana condition on the fermions, and more especially, how the
self-duality condition on the 5-form, is to be handled.  In fact the
self-duality condition will force the 5-form to be complex, implying
that the T-dual type IIA Euclidean-signature theory will also be
complex.  Thus T-duality would have to be discussed in situations
where some of the relevant solutions, and indeed the theories
themselves, are complex.

    By contrast, our prescription is simply to diagonally reduce a
$p$-brane until it becomes an instanton, perform an $SL(2,\R)$ duality
transformations to shift and rescale the harmonic function, and then
diagonally oxidise back to the $p$-brane again.  $N$-charge
$p$-branes or intersections (\ie with $N$ independent harmonic
functions) are handled similarly, by reducing down to the dimension
where they become $N$-charge instantons, which are described by a
Euclidean-signature theory with $N$ independent $SL(2,\R)$ factors in
the the global symmetry group that allow the $N$ harmonic functions to
be independently shifted and scaled.  Although the instantons are
described by Euclidean-signature theories, these arise naturally from
reduction on the time coordinate, and no act of Euclideanisation is
performed.  Also, our discussion extends to the case of non-extremal
solutions, which would not be possible in the D-instanton approach
described in \cite{bb}, since type IIB supergravity does not have the
$GL(2,\R)/O(1,1)$ scalar manifold that would be needed for
constructing non-extremal instantons.

\section{$SL(3,\R)/O(2,1)$ Lagrangians, and instantons}

    In this section, we give some explicit results for the
$SL(3,\R)$-symmetric part of the scalar Lagrangian for
eight-dimensional Euclidean-signature supergravity.  This is a useful
example because it is exhibits more ``generic'' behaviour than is seen
in the $SL(2,\R)$ example in type IIB or in $D=9$.  In particular,
there are three axions (plus two dilatons) involved in the $SL(3,\R)$
scalar manifold, and the axions can undergo rotations under
$SL(3,\R)$, for which there is no analogue in the single-axion
$SL(2,\R)$ system, in addition to non-linear transformations of a kind
that are familiar in $SL(2,\R)$.  At the same time, $SL(3,\R)$ is
still sufficiently simple that explicit formulae can be presented.  We
shall present results for the case where the eight-dimensional space
is of Euclidean signature.  We shall shall obtain this theory by
taking the time reduction to be at the $D=10$ to $D=9$ stage of the
dimensional reduction process.

\subsection{$SL(3,\R)$-invariant scalar Lagrangians}

     The Lagrangian can be obtained from the general results in
section 2.1, with the sign reversal occurring for the kinetic terms of
all field strengths carrying the index value ``2.''  .  Thus we find
that the relevant part of the Lagrangian is
\be
e^{-1}\, {\cal L} = -\ft12 (\del\vec\phi)^2 -\ft12 e^{\vec
b_{13}\cdot\vec\phi}\, (\del \chi_{13}- \chi_{23}\, \del\chi_{12})^2 
+\ft12 e^{\vec b_{12}\cdot\vec\phi}\, (\del\chi_{12})^2 
+\ft12 e^{\vec b_{23}\cdot\vec\phi}\, (\del\chi_{23})^2\ .
\label{sl3rlag}
\ee
We have suppressed the extra $SL(2,\R)$ invariant part of the full
$D=8$ scalar Lagrangian.  This comprises a 1-dilaton, 1-axion system.
In fact in the full scalar Lagrangian in $D=8$, there will be a
3-vector $\vec\phi$ of dilatons, and the extra axion $\chi_{123}$ of the
$SL(2,\R)$ system, coming from the dimensional reduction of $A_\3$ in
$D=11$.  Its dilaton coupling in the Lagrangian, $\ft12 e^{\vec
a_{123}\cdot\vec\phi}\, (\del\chi_{123})^2$, involves a dilaton vector
$\vec a_{123}$ that is orthogonal to all three of the $\vec b_{ij}$
dilaton vectors in (\ref{sl3rlag}), and in writing the pure
$SL(3,\R)$ system in (\ref{sl3rlag}), we are taking $\vec\phi$ to be
just a 2-component vector of dilatons in the directions orthogonal to
$\vec a_{123}$.  In fact in this basis, we are taking the dilaton
vectors $\vec b_{ij}$ to be
\be
\vec b_{12} =(\sqrt3, -1)\ ,\qquad
\vec b_{23} =(-\sqrt3, -1)\ ,\qquad
\vec b_{13}= (0,-2)\ .\label{sl3b}
\ee

     Following \cite{cjlp}, we can parameterise an $SL(3,\R)/O(2,1)$ 
coset representative $\v$, in the Borel gauge, as
\bea
\v &=& e^{\fft12 \vec\phi\cdot\vec H}\, e^{\chi_{23}\, E_{23}}\,
 e^{\chi_{13}\, E_{13}}\,  e^{\chi_{12}\, E_{12}}\ ,\nn\\
&&\nn\\
&=& \pmatrix{ e^{\fft1{2\sqrt3}\phi_1 -\fft12\phi_2} &
    \chi_{12}\, e^{\fft1{2\sqrt3}\phi_1 -\fft12\phi_2} &
 \chi_{13}\, e^{\fft1{2\sqrt3}\phi_1 -\fft12\phi_2} \cr
0 & e^{-\fft1{\sqrt3}\phi_1} & \chi_{23}\, e^{-\fft1{\sqrt3}\phi_1} \cr
0 & 0 & e^{\fft1{2\sqrt3}\phi_1 +\fft12\phi_2} } \ ,\label{sl3v}
\eea
where $\vec H$ represents the two Cartan generators, and $E_{ij}$
denote the positive-root generators of $SL(3,\R)$.  Defining
\be
{\cal M} = \v^T\, \eta\, \v\ ,\qquad\qquad
\eta={\rm diag}\, (1,-1,1)\ ,\label{mvv}
\ee
the Lagrangian (\ref{sl3rlag}) can be written as
\be
e^{-1}\, {\cal L} = \ft14{\rm tr}\, (\del_\mu {\cal M}^{-1}\, \del^\mu{\cal
M})\ .\label{sl3rlag2}
\ee

     Global $SL(3,\R)$ transformations on the scalar fields can be
implemented by acting on the right of $\v$ with a constant $SL(3,\R)$ matrix
$\Lambda$, and on the left with a local compensating $O(2,1)$
transformation ${\cal O}$, whose job is to restore the transformed $\v$
to the Borel gauge:
\be
\v \longrightarrow \v' = {\cal O}\, \v \, \Lambda\ .\label{trans}
\ee
It is manifest that provided ${\cal O}$ satisfies ${\cal O}^T\,\eta\,
{\cal O}=\eta$, this will leave the Lagrangian (\ref{sl3rlag2})
invariant for any global $SL(3,\R)$ transformation matrix $\Lambda$.   

     It is of particular interest to study the transformations of the 
scalar fields under the $O(2,1)$ subgroup of $SL(3,\R)$, since this is
the subgroup that preserves a given set of values for the scalars.
Thus we may choose the particular $O(2,1)$ subgroup that preserves the
values of the scalar moduli, \ie the asymptotic values of the scalar
fields at infinity.  The simplest choice is to take all the moduli to
be zero, since then the coset representative (\ref{sl3v}) is simply
the identity, and so the required $O(2,1)$ subgroup will consist just
of matrices $\Lambda$ that satisfy
\be
\Lambda^T\, \eta\, \Lambda = \eta\ .\label{o21sub}
\ee
It is somewhat involved, even in this special case, to give a
parameterisation of all such $O(2,1)$ matrices, and the resulting
expressions for the transformed scalar fields will be quite
complicated.  However, it suffices that we derive the transformations
for two different 1-parameter subgroups, namely the $O(2)$ subgroup of
matrices
\be
\Lambda_1 = \pmatrix{\cos\theta & 0 & \sin\theta\cr
                     0 & 1 & 0 \cr
                     -\sin\theta & 0 &\cos\theta}\ ,\label{o2matrix}
\ee
and the $O(1,1)$ subgroup of matrices
\be
\Lambda_2 = \pmatrix{\cosh t & \sinh t & 0\cr
                     \sinh t & \cosh t & 0\cr
                      0 & 0 & 1}\ .\label{o11matrix}
\ee
Any desired $O(2,1)$ transformation can be obtained by composing these
two basic transformations. 

     In each case, we may obtain the transformation rules for the
scalar fields by substituting $\Lambda_1$ or $\Lambda_2$ into
(\ref{trans}), solving for the compensator ${\cal O}$ that restores
the Borel gauge, and then reading off the transformed fields from the
resulting Borel matrix $\v'$.  For the $O(2)$ transformation
$\Lambda_1$ given in (\ref{o2matrix}), it is useful first to define
the function
\be
f_1= e^{\sqrt3 \phi_1}\, (\cos\theta - \chi_{13}\, \sin\theta)^2 +
      e^{\sqrt3\phi_1+2\phi_2}\, \sin^2\theta 
        -\chi_{23}^2\, e^{\phi_2}\, \sin^2\theta\ .\label{fdef}
\ee
We find that the $O(2)$ transformations then take the form
\bea
e^{-2\phi_1'/\sqrt3} &=&  f_1^{-1}\, e^{\phi_1/\sqrt3}
\Big((\cos\theta -(\chi_{13}-\chi_{12}\, \chi_{23})\sin\theta)^2 
 +e^{2\phi_2}\, \sin^2\theta \nn\\
&&\qquad\qquad\qquad -\chi_{12}^2\, e^{\sqrt3\phi_1+\phi_2}
 \sin^2\theta)\Big) \ ,\nn\\
e^{\phi_1'/\sqrt3 -\phi_2'} &=& f_1\, e^{-2\phi_1/\sqrt3 -\phi_2}\ ,\nn\\
e^{\phi_1'/\sqrt3 -\phi_2'}\, \chi_{12}' &=& \chi_{12}\, e^{\phi_1/\sqrt3
-\phi_2}\, (\cos\theta -\chi_{13}\, \sin\theta)
       + \chi_{23}\, e^{-2\phi_1/\sqrt3}\, \sin\theta \ ,\nn\\
e^{\phi_1'/\sqrt3 -\phi_2'}\, \chi_{13}' &=& 
(\chi_{13}\cos\theta +\sin\theta)(\cos\theta-\chi_{13}\sin\theta)\, 
e^{\phi_2/\sqrt3 -\phi_2}  \label{o2rot}\\
&& \qquad\qquad -e^{\phi_2/\sqrt3 +\phi_2}\sin\theta\cos\theta
+ \chi_{23}^2\, e^{-2\phi_1/\sqrt3}\, \sin\theta\cos\theta \ ,\nn\\
e^{-2\phi_1'/\sqrt3}\, \chi_{23}' &=& f_1^{-1}\, e^{\phi_1/\sqrt3}\,
\Big(\chi_{23}(\cos\theta-(\chi_{13} -\chi_{12}\, \chi_{23})\sin\theta)
-\chi_{12}\, e^{\sqrt3\phi_1 +\phi_2}\, \sin\theta\Big)\ .\nn
\eea
For the $O(1,1)$ transformations given by $\Lambda_2$ in
(\ref{o11matrix}), we define 
\be
f_2 = (\cosh t + \chi_{12}\, \sinh t)^2\, e^{\sqrt3\phi_1} -
e^{\phi_2}\, \sinh^2 t\ ,
\ee
and we find that 
\bea
e^{-2\phi_1'/\sqrt3} &=& f_2^{-1}\, e^{\phi_1/\sqrt3} \ , \nn\\
e^{\phi_1'/\sqrt3 -\phi_2'} &=& f_2\, e^{-2\phi_1/\sqrt3 -\phi_2}\ , \nn\\
e^{\phi_1'/\sqrt3 -\phi_2'}\, \chi_{12}' &=& (\chi_{12}\, \cosh t +
\sinh t)(\cosh t + \chi_{12}\, \sinh t)\, e^{\phi_1/\sqrt3 -\phi_2}\nn\\
&& \qquad -e^{_-2\phi_1/\sqrt3}\, \sinh t\cosh t \ ,\nn\\
e^{\phi_1'/\sqrt3 -\phi_2'}\, \chi_{13}' &=& \chi_{13}\, 
(\cosh t + \chi_{12}\, \sinh t)\, e^{\phi_1/\sqrt3 -\phi_2} 
 -\chi_{23}\, e^{-2\phi_1/\sqrt3}\, \sinh t \ ,\nn\\
e^{-2\phi_1'/\sqrt3}\, \chi_{23}' &=& f_2^{-1}\,e^{\phi_1/\sqrt3}\,
(\chi_{23}\, \cosh t -(\chi_{13}-\chi_{12}\, \chi_{23})\, \sinh t)\ .
\label{o11rot}
\eea

     Let us now turn to an explicit demonstration for the
$SL(3,\R)/O(2,1)$ scalar manifold of the claim that we made in section
2.2, that the order in which the time and the space reductions are
performed does not affect the final form of the lower-dimensional
Lagrangian.  In particular, we shall show that the Lagrangian
(\ref{sl3rlag}) obtained by reducing on $t$ at the second step is the
same as the Lagrangian obtained by reducing on $t$ instead at the
first step.  This latter Lagrangian is
\be
e^{-1}\, \widetilde{\cal L} = -\ft12 (\del\vec\phi')^2 +\ft12 e^{\vec
b_{13}\cdot\vec\phi'}\, (\del \chi_{13}'- \chi_{23}'\, \del\chi_{12}')^2
+\ft12 e^{\vec b_{12}\cdot\vec\phi'}\, (\del\chi_{12}')^2
-\ft12 e^{\vec b_{23}\cdot\vec\phi'}\, (\del\chi_{23}')^2\ ,
\label{sl3rlag3}
\ee
in terms of a primed set of field variables.  At first sight, it is
far from obvious that this is equivalent to (\ref{sl3rlag}),
especially in view of the fact that the ``distinguished'' axion
$\chi_{13}$ whose field strength receives the Kaluza-Klein
modification has a kinetic term with opposite signs in the two
cases. To show that in fact the Lagrangians are the same, but written
in different field variables, we shall give a slightly different proof
from the general one that we presented in section 2.2.  In particular,
we shall derive an explicit purely real field transformation here (\ie
real within a neighbourhood).  To do this, we note that just as
(\ref{sl3rlag}) can be written as ${\cal L}= \ft14 e\, {\rm
tr}(\del_\mu{\cal M}^{-1}\, \del^\mu {\cal M})$, where ${\cal M}=\v^T\, \eta\,
\v$, and $\eta ={\rm diag}(1,-1,1)$, so (\ref{sl3rlag3}) can be
written as $\widetilde {\cal L}= \ft14 e\, {\rm tr}(\del_\mu({\cal
\widetilde M}')^{-1}\, \del^\mu {\cal \widetilde M'})$, where $\widetilde
{\cal M'}= {\v'}^T\, \td\eta\, \v'$, and $\td\eta={\rm diag}(-1,1,1)$.
Now let us consider an $SL(3,\R)$ transformation $\Lambda$, and define
a transformed Borel-gauge coset representative by $\v' = {\cal C}\,
\v\, \Lambda$, where the ``compensating transformation'' ${\cal C}$ is
required to satisfy
\be
{\cal C}^T\, \td\eta\, {\cal C} = \eta\ .
\ee
Then we find that the Lagrangian (\ref{sl3rlag3}), \ie 
$\widetilde {\cal L}= \ft14 e\, {\rm tr}(\del_\mu({\cal \widetilde M'})^{-1}\,
\del^\mu {\cal \widetilde M'})$, is mapped by this transformation into
the Lagrangian (\ref{sl3rlag}), expressed in terms of the unprimed
fields. Taking the $SL(3,\R)$ transformation to be
\be
\Lambda = \pmatrix{ 0&1&0\cr
                    -1&0&0\cr
                    0&0&1}\ ,  \label{sl3rlam}
\ee  
we find that the explicit transformation between the primed fields in
(\ref{sl3rlag3}) and the unprimed fields in (\ref{sl3rlag}) is
\bea
\phi_1' &=& -\ft12 \phi_1 +\fft{\sqrt3}{2}\, \log f\ ,\nn\\
\phi_2' &=& \fft{\sqrt3}{2} \phi_1 + \phi_2 -\ft12 \log f\ ,\nn\\
\chi_{12}' &=& f^{-1}\, e^{\sqrt3\phi_1}\, \chi_{12}\ ,\label{transf}\\
\chi_{13}' &=& -\chi_{23}\, f^{-1}\, e^{\phi_2} + f^{-1}\, \chi_{12}\, 
\chi_{13}\, e^{\sqrt3\phi_1}\ ,\nn\\
\chi_{23}' &=& \chi_{13} - \chi_{12}\, \chi_{23}\ ,\nn 
\eea
where 
\be
f = e^{\phi_2} - e^{\sqrt3 \phi_1}\, \chi_{12}^2\ .
\ee
Note that this transformation is real in the patch where $f>0$.  This
is an example of the general result that we described in section 2.2,
where the transformation between the ${\cal M}$ matrices
parameterising the cosets in the two equivalent Lagrangians can be
arranged to induce a real transformation between the two sets of coset
coordinates, by appropriate choice of parameterisation.  In the
notation of section 2.2, our example here corresponds to taking ${\cal
M}(2) = \Lambda\, {\cal M}'(1)\, \Lambda^{-1}$, with $\Lambda$ given
by (\ref{sl3rlam}).

\subsection{$SL(3,\R)$ transformation of instantons}

       Having studied the full global $SL(3,\R)$ symmetry of the
Lagrangian (\ref{sl3rlag}) for the coset $SL(3,\R)/O(2,1)$, we are in
a position to investigate how the instanton solutions transform under
the $SL(3,\R)$ global symmetry.  There are a total of three axions,
namely $\chi_{12}$, $\chi_{23}$ and $\chi_{13}$, each of which can
support a simple single-charge instanton of the form
\bea
ds^2 &=& dr^2 + r^2 \d\Omega^2\ ,\nn\\
\vec \phi &=& \ft12\vec b_{ij} \log H\ ,\qquad \chi_{ij} = H^{-1}
\ .\label{sl3rsol}
\eea
Note that for the instantons supported by either $\chi_{12}$ or
$\chi_{23}$, the charge $Q$ in the harmonic
function $H=1 + Q r^{-\td d}$ is real since these axions are NC-scalars.
For $\chi_{13}$, on the other hand, the charge $Q$ is imaginary.  The 
scalar coset matrices ${\cal M}$ for these three solutions are given by
\bea
\chi_{12}=H^{-1}:&& {\cal M} = \pmatrix{H & 1 & 0\cr 1 & 0 & 0\cr
                                        0 & 0 & 1}\label{chi12sol}\\
&&\nn\\
\chi_{23}=H^{-1}:&& {\cal M} = \pmatrix{1 & 0 & 0 \cr 0 & -H & -1 \cr
                                       0 & -1 & 0}\label{chi23sol}\\
&&\nn\\
\chi_{13}=H^{-1}:&& {\cal M} =\pmatrix{H & 0 & 1\cr 0 & -1 & 0\cr
                                       1 & 0 & 2H^{-1}}\label{chi13sol}
\eea
Thus it is straightforward derive how the scalar fields in these three
solutions transform under the $SL(3,\R)$ transformation ${\cal M}
\rightarrow \Lambda^{\rm T} {\cal M} \Lambda$.

     We shall now examine the transformations of the Noether
currents for the instanton solutions under
the $SL(3,\R)$ symmetry.  In Appendix A, we present results for the eight
Noether currents associated with the parameters of the global
$SL(3,\R)$ symmetry.  These are the analogues of the three $SL(2,R)$
Noether currents given in (\ref{sl2rcurrents}).  We also show that
these transform linearly under $SL(3,\R)$. 

    It is a straightforward matter to
substitute the above instanton solutions, but with the scalar moduli chosen 
to be zero for simplicity, into the set of eight Noether currents
given in (\ref{sl3rnoether}) in Appendix A.  (The solutions with zero
moduli are obtained from those given in (\ref{sl3rsol}) by performing
a shift Borel transformation so that now $\chi_{ij}=H^{-1} -1$, with
all other fields unchanged.)  We find that the
instanton supported by the axion $\chi_{12}$ has Noether currents
given by
\be
{\cal J}(\chi_{12})=  \pmatrix{-dH & dH & 0\cr
                   -dH & dH & 0\cr
                    0 & 0 & 0}\ .\label{q12charge} 
\ee
The instanton supported instead by $\chi_{23}$ has Noether currents given by
\be
{\cal J}(\chi_{23}) = \pmatrix{0 & 0 & 0\cr
                     0 & -dH & dH \cr
                    0 & -dH & dH}\ .\label{q23charge}
\ee
Finally, the Noether currents for the complex solution supported by
$\chi_{13}$ are given by
\be
{\cal J}(\chi_{13})  = 
\pmatrix{ (1-2H^{-1}) dH & 0 & (-1+4H^{-1} -2H^{-2})dH\cr
                      0 & 0 & 0\cr
                   dH & 0 & (-1+ 2H^{-1}) dH}\ .
\ee
In each case, $dH$ is the exterior derivative of the harmonic 
function $H$ characterising the solution.

     It is easy to verify that the $O(2)$ transformation $\Lambda_1$
given in (\ref{o2matrix}) rotates the two sets of Noether currents for
the real solutions using $\chi_{12}$ and $\chi_{23}$ into one another,
and in fact if the rotation parameter $\theta$ is chosen to be
$3\pi/2$, then we find that $\Lambda_1^{-1}\, {\cal J}(\chi_{12})\,
\Lambda_1 = {\cal J}(\chi_{23})$.  It is also evident that the
$O(1,1)$ transformation (\ref{o11matrix}) acts on the $\chi_{12}$
solution in the same way as did the $O(1,1)$ transformation
(\ref{sl2ro11}) in the $SL(2,\R)$ instanton solution discussed in
section 5.1, namely as an overall rescaling of the Noether
currents. The other $O(1,1)$ subgroup transformation that we did not
write down, corresponding to a Lorentz rotation in the 2-3 plane
rather than the 1-2 plane of the transformation (\ref{o11matrix}),
would act similarly on the $\chi_{23}$ instanton solution.

    Thus we see that the orbits of the modulus-preserving $O(2,1)$
subgroup of the $SL(3,\R)$ symmetry group of the scalar Lagrangian
(\ref{sl3rlag}) include an $O(2)$ subgroup that rotates between the
pair of instanton solutions supported by $\chi_{12}$ and by
$\chi_{23}$.  It is important to emphasise, however, that we have only
a doublet, and not a triplet, of instanton solutions in this example,
despite the occurrence of three axions in the Lagrangian.  The reason
for this is that only two out of the three axions have kinetic terms
with the necessary sign to allow them to support real instanton
solutions, and only these two can rotate into one another under the
action of the real global symmetry transformations.

\section{$(D-3)$-branes}

     Our principal focus so far in this paper has been on the
investigation of  Euclidean-signature maximal supergravities, and the
real instanton solutions that can be supported by those axions whose
kinetic terms have undergone a sign reversal in the reductions to the
Euclidean-signature theories.  

      Axions can also support $(D-3)$-brane solitons in
$D$-dimensional Minkowskian-signature spacetimes.  One might think
that they can simply be viewed as the magnetic duals of the
instantons, but we shall shortly see that their relationship is more
complicated than that.  There are different types of
$(D-3)$-branes. First let us consider the ones that can be viewed as
coming from the vertical dimensional reduction of standard $p$-branes,
until the point is reached where the transverse space becomes
two-dimensional.  Such solutions have the following structure
\bea
ds^2&=& dx^\mu dx_\mu + (1 + \ft{Q}{2\pi}\log r) 
(dr^2 + r^2d\theta^2)\ ,\nn\\
e^{-\phi}& = &1 + \ft{Q}{2\pi}\log r\ ,\qquad
\chi = \ft{Q}{2\pi} \theta\ . \label{sl2rdm3sol}
\eea
It was argued in \cite{sen} in the context of strings in
four-dimensional theories that such solutions break the classical
$SL(2,\R)$ duality symmetry down to the quantum S-duality group
$SL(2,\Z)$, since the periodicity of the angular coordinate $\theta
=\theta + 2\pi$ implies that $\chi$ also must become periodic, with
$\chi =\chi +1$ in the case of a string carrying a unit charge $Q$.
The magnetic charge of the solution can be defined as $Q_m=\int J_m$,
where $J_m=d\chi$ is the current dual to $e^{2\phi}\,
{*d\chi}={*J_+}$, whose integral gives the electric charge of the
instanton. It is conserved, by the virtue of the Bianchi identity
$dJ_m=0$.  However, $J_m$ is not invariant under $SL(2,\R)$, and in
fact acting on $J_m$ with $SL(2,\R)$ generates an infinite number of
currents
\bea
J_m^1 &=& d\chi\ ,\nn\\
J_m^2 &=& \chi d\chi + e^{-2\phi} d\phi\ ,\nn\\
J_m^3 &=& \chi^2 d\chi + 2e^{-2\phi} \chi d\phi -e^{-2\phi} d\chi
\ ,\label{dualcurrents}\\
J_m^4 &=& \chi^3 d\chi + 3e^{-2\phi} \chi^2 d\phi -
3e^{-2\phi} \chi d\chi - e^{-4\phi} d\phi\ .\nn\\
&&\cdots\cdots\nn
\eea
(An analogous calculation of the action of $SL(2,\R)$ on the Noether
current $J_+$ just gives the closed system of three Noether currents
$J_+$, $J_0$ and $J_-$.)  The currents (\ref{dualcurrents}) form an
infinite-dimensional representation of $SL(2,\R)$.  The $(D-3)$-brane
solution evidently cannot be viewed as the dual of the instanton
solution discussed earlier, since in that case the Noether charges are
in the adjoint representation of $SL(2,\R)$.  In fact the manifest
occurrence of bare undifferentiated $\chi$ fields in
(\ref{dualcurrents}) implies that the charges of the $(D-3)$-brane,
calculated from (\ref{dualcurrents}), except for $\int J_m^1$ itself,
are ill-defined, since $\chi = \ft{Q}{2\pi}\, \theta$ is not a
periodic function of $\theta$ in the solution.  This manifests itself
in the fact that quantities such as $\oint \theta\, d\theta$ are
ill-defined.  In fact the $(D-3)$-brane solution breaks the $SL(2,\R)$
symmetry group down to the group of integer-valued strict Borel
transformations,
\be
 \Lambda = \pmatrix{1 & n\cr
                    0 & 1}\ .
\ee

     There also exists an $SL(2,\R)$-invariant $(D-3)$-brane
\cite{gsvy,ggp}. It might be this solution that is dual to the
instanton we discussed earlier.  Athough the solution
(\ref{sl2rdm3sol}) may be incompatible with the U-duality, it provides
a starting-point for obtaining a domain-wall solution of a massive
supergravity in one dimension lower.  This massive supergravity, which
does not inherit the full U-duality from the higher dimension, is
obtained by making a Scherk-Schwarz reduction on the $\theta$
coordinate.  And indeed, the domain-wall solution has no magnetic
dual.

     In the rest of this section, we shall consider in some detail a
subset of the set of all $(D-3)$-brane solitons for which the
complications described above can be avoided.  Specifically, we shall
consider exclusively $(D-3)$ branes that are supported by axions
coming from the dimensional reduction of the $A_\3$ potential of
eleven-dimensional supergravity.  Furthermore, we shall consider the
action on such solutions of only the $SL(11-D,\R)$ subgroups of the
full global supergravity symmetry groups, which we shall call the
``restricted symmetry groups.''  Consequently, we shall be considering
axions that undergo only linear transformations under the restricted
global symmetry group.

     The method that we shall use in order to study the $(D-3)$-brane
multiplets is analogous to the one used in \cite{lpsorbit} for studying
the multiplet structures for $p$-branes supported by higher-degree
field strengths.  Here, we are concerned only with the global
$SL(11-D,\R)$ symmetry, and the associated positive-root generators
$E_i{}^j$.  Since we are considering only $(D-3)$-branes that are
supported by axions derived from the potential $A_\3$ of $D=11$
supergravity, the highest dimension in which any such solution can
exist is $D=8$.  Furthermore, since there is only one axion of this
kind in $D=8$, namely $A_{\0 123}$, the associated 5-brane is a
singlet.  Thus we must descend to $D=7$ before encountering an
interesting multiplet structure.

     In $D=7$ the restricted symmetry group is $SL(4,\R)$, with simple
roots $\vec{b}_{i,i+1}$ where $i=1,2,3$. The full root system is given
by $\pm\vec{b}_{ij}$, where $i,j=1,2,3,4$, associated with the
generators $E_{\pm\vec{b}_{ij}}$.  In the standard basis for
$SL(n,\R)$, the Cartan generators can be written as
$H_i=E_i{}^i-E_{i+1}{}^{i+1}$.  We wish to consider 4-brane solutions
supported by the 1-form field strengths $F_{\1 ijk}$.  These fields
form a 4-dimensional representation of the $SL(4,\R)$ algebra.  We can
determine the orbits of the 4-brane solutions by picking a
representative solution, and considering the stability subgroup ${\cal
H}$ of $SL(4,\R)$ that leaves the solution invariant.  The orbits will
then be given by the coset $SL(4,\R)/{\cal H}$. Let us, for
definiteness, pick the solution supported by $F_{\1 123}$ as our
starting point.  The dilaton vector $\vec a_{123}$ for this field is
the highest-weight vector in the 4-dimensional representation. Thus we
now need to find the subset of $SL(4,\R)$ generators that annihilate
the highest-weight state $|\vec a_{123} \rangle$.  It is
straightforward to check that from the three Cartan generators there
are just two combinations that annihilate this state, namely
$H_1=E_1{}^1-E_2{}^2$, and $H_2=E_2{}^2-E_3{}^3$. Of the remaining
$SL(4,\R)$ generators, the following annihilate $|\vec a_{123}
\rangle$:
\be
E_1{}^{2},\quad E_2{}^{1}, \quad E_1{}^{3}, \quad E_3{}^{1}, \quad E_2{}^{3},
\quad E_3{}^{2}, \quad E_4{}^{1}, \quad E_4{}^{2}, \quad E_4{}^3
\la{gd7}
\ee
Under the two Cartan combinations $(H_1,H_2)$ , the
weights of the first six generators in (\ref{gd7}), which form three
conjugate pairs $\{E_i{}^j, E_j{}^i\}$, are
\be
(2, -1);\ (-2, 1);\ (1, 1);\ (-1, -1);\ (-1, 2);\
(1, -2)\ .
\la{rd7}
\ee From the new Cartan generators one can construct the Killing metric
\be 
g_{ij}= \tr(H_i\, H_j) = \left(\begin{array}{cc}
                          2  & -1\\
                          -1 &  2\end{array}\right)\ .
\la{kd7}
\ee
Defining the sign of a root by the sign of its first non-zero
component found working in from the left, one can easily see 
that the simple roots are $\a_1=(1, 1)$ and $\a_2=(1, -2)$.  Their dot
products, defined using the Killing metric (\ref{kd7}), are given by
\be
\a_1\cdot \a_1=2\ ,\quad \a_2\cdot\a_2 =2\ ,\quad \a_1\cdot\a_2=-1\ .
\la{sd7}
\ee
We therefore see that $E_1{}^{2}, E_2{}^{1}, E_1{}^{3}, E_3{}^{1},
E_2{}^{3}$ and $E_3{}^{2}$, together with $H_1$ and $H_2$, generate an
$SL(3,\R)$ algebra.  Furthermore, the remaining three generators
$E_4{}^{1}, E_4{}^{2}$ and $E_3{}^4$ mutually commute, and form a vector
representation under $SL(3,\R)$.  Thus the coset space parameterising
the single-charge 1-form solutions in $D=7$ is
\be
\frac{SL(4,\R)}{SL(3,\R)\semi \R^3}\ .
\la{csd7}
\ee

     We shall not present the analogous detailed calculations in lower
dimensions, and instead we shall just give the results. The cosets
describing the orbits under the restricted symmetry groups for
single-charge $(D-3)$-branes are listed in Table 6.

\bigskip\bigskip

\begin{center}
\begin{tabular}{|c|c|}\hline
 & Coset  \\ \hline
$D=7$ & $\fft{SL(4,\sR)}{SL(3,\sR)\semi \sR^3}$  \\ \hline
$D=6$ & $\fft{SL(5,\sR)}{SL(3,\sR)\times SL(2,\sR)\semi \sR^6}$  \\ \hline
$D=5$ & $\fft{SL(6,\sR)}{SL(3,\sR)\times SL(3,\sR)\semi \sR^6}$  \\ \hline
$D=4$ & $\fft{SL(7,\sR)}{SL(3,\sR)\times SL(4,\sR)\semi \sR^{12}}$  \\ \hline
$D=3$ & $\fft{SL(8,\sR)}{SL(3,\sR)\times SL(5,\sR)\semi \sR^{15}}$  \\ \hline
\end{tabular}
\end{center}

\bigskip

\centerline{Table 6: Cosets for single-charge $(D-3)$-brane orbits}

\bigskip\bigskip

    In addition to these single-charge $(D-3)$-branes, there are also, in
lower dimensions, multi-charge $(D-3)$-brane solutions for which the
natures of the orbits are different.  We shall just present one example
here, to illustrate the procedure.  The simplest example
that illustrates the point occurs in $D=6$.  We see from Table 6 that
the single-charge 3-brane solution has orbits of dimension 7, while
the number of 1-form field strengths $F_{\1 ijk}$ is 10.  (By
contrast, in $D=7$ the orbits have dimension 4, which is equal to the
number of field strengths $F_{\1 ijk}$.)  The fact that in $D=6$ the
single-charge orbits have a smaller dimension than the number of
available field strengths that could support the solutions suggests
that there should exist new classes of solution, that would ``fill
out'' orbits of higher dimension. Indeed, in $D=6$ the possibility
arises for the first time of having 2-charge 3-brane solutions,
carrying two independent charges.  An example is a solution whose two
charges are carried by the field strengths $F_{\1 123}$ and $F_{\1 145}$.
The orbits of this solution can then be determined by the same methods
as above, namely by first identifying the stability group that leaves
both of the associated root vectors $\vec a_{123}$ and $\vec a_{145}$
simultaneously invariant.  This turns out to be $Sp(4)\semi \R^4$.  
Thus the coset describing the 2-charge orbits is
\be
\fft{SL(5,\R)}{Sp(4)\semi \R^4}\ .
\ee
This coset has dimension 10, and so one can expect that the orbits for
these solutions indeed fill out the entire solution space.  And indeed,
there do not exist any more general 3-charge solutions in $D=6$.

\section{Discussion}

    In this paper, we have obtained the Euclidean-signature
supergravities that result from compactifying $D=11$ supergravity or
type IIB supergravity on a torus that includes the time direction.
These Euclidean-signature theories are automatically compatible with
any Majorana or self-duallity conditions on fields, since they are
obtained by a consistent dimensional-reduction procedure.  We showed
that there are two inequivalent nine-dimensional theories, coming from
the reduction of the type IIA and type IIB supergravities on their
time directions.  The two nine-dimensional Euclidean-signature
theories become equivalent upon further compactification on a spatial
circle.  This can also be understood from the general result that the
same Euclidean-signature theory is obtained regardless of the order in
which the time reduction and spatial reductions are performed.  We
studied the global symmetry groups of the Euclidean-signature
theories, and the structure of their scalar cosets.  We also
investigated the orbits of instanton solutions under the global
symmetry groups in the examples of $SL(2,\R)$ and $SL(3,\R)$-invariant
Lagrangians.

   We showed that the $SL(2,\R)$ symmetry of the Euclidean-signature
theory which describes the instanton coming from the diagonal
dimensional reduction of a $p$-brane on its entire
world-volume\footnote{It has recently been argued that it is necessary
to consider the wrapping of $p$-branes on the time as well as spatial
world-volume directions in a full discussion of their singularity
structure \cite{gibbons}.} can transform the $p$-brane into its
near-horizon structure.  In the case of non-dilatonic $p$-branes the
curvature, and the singularity structure of the $p$-brane, can be
completely different from its near-horizon behaviour.  For example,
the eleven-dimensional membrane \cite{dust} has a curvature
singularity, and it requires the inclusion of the membrane action
\cite{bst} as a source term \cite{dkl}.  On the other hand, its
near-horizon structure is AdS$_4\times S^7$, which is an exact
supergravity solution without any singularity and with no need for a
source term.  This emphasises that the lower-dimensional U-duality
groups must be more than just the residues of the general-coordinate
symmetries and gauge symmetries of the eleven-dimensional
theory.\footnote{For example, it is known that the global homogeneous
scaling transformation of the eleven-dimensional theory plays an
essential r\^ole in the global symmetry transformations in $D\le 10$
\cite{cjlp}.}  For example, the eleven-dimensional membrane becomes an
instanton in $D=8$, after it is reduced on its 3-dimensional
world-volume.  The instanton is supported by the axion $A_{\0 123}$,
coming from the reduction of $A_\3$ in $D=11$, and by a dilaton $\phi=
\ft12 \vec a_{123} \cdot \vec\phi$, which comes from the metric.  The
$SL(2,\R)$ symmetry of this system, which we used in order to
transform the structure of the harmonic function, is the $SL(2,\R)$
factor of the $SL(3,\R)\times SL(2,\R)$ U-duality group, and it is
therefore distinct from the $SL(3,\R)$ which comes from the general
coordinate symmetry of the 3-torus.  Note that this $SL(2,\R)$
symmetry is part of the T-duality symmetry $O(2,2)$ of the type IIA
string in $D=8$. In $D=11$, it is a symmetry that mixes the metric
and the 3-form gauge potential.

    On the other hand, in cases where the $SL(2,R)$ symmetry of the
theory describing the instanton does come from the general coordinate
symmetry in the internal space, the constant shift of the harmonic
function in the higher-dimensional solution will not affect its
curvature.  For example, the D0-brane in the type IIA theory, which
can be viewed as a wave in $D=11$, can be reduced on its time
direction to an instanton in $D=9$.  The $SL(2,\R)$ symmetry in $D=9$
is just the general-coordinate symmetry of the internal torus, and so
in this case there should be no change in the curvature or singularity
structure.  Indeed, any constant shift or rescaling of the harmonic
function of the wave solution can be achieved by a general-coordinate
transformation \cite{hyun,bps}.

     The fact that the U-duality groups in $D\le8$ dimensions can
alter the singularity structure of M-branes suggests that a better
understanding of U-duality from the higher-dimensional viewpoint is
needed.

\appendix
\section{$SL(3,\R)$ Noether currents}

    In this appendix, we present the detailed expressions for the
eight Noether currents corresponding to the eight parameters of the 
global $SL(3,\R)$ symmetry of the scalar Lagrangian (\ref{sl3rlag}).
In fact it is convenient first to present a more general derivation of
the Noether currents for an arbitrary scalar Lagrangian of the form
\be
{\cal L} = \ft14 e\, {\rm tr}\, (\del_\mu{\cal M}^{-1}\,\del^\mu 
{\cal M})\ .\label{mlag}
\ee
This is invariant under the global $G$ transformations
\be
{\cal M}\longrightarrow {\cal M}' = \Lambda^T\, {\cal M}\, \Lambda\ .
\label{globaltrans}
\ee
Infinitesimally, where we write $\Lambda=\oneone + \lambda$, and
$\lambda$ is infinitesimal, we have
\be
\delta{\cal M} = \lambda^T\, {\cal M} + {\cal M}\, \lambda\ .\label{inftrans}
\ee
By the usual procedure, we calculate the Noether currents by varying
the Lagrangian with respect to a spacetime-dependent transformation,
keeping only those terms where a derivative falls on the parameters
$\lambda$.  Thus we have
\bea
\delta{\cal L} &=& -\ft12 e\, \Big({\cal M}^{-1}(\del_\mu \lambda^T\, {\cal
M} +{\cal M}\, \del_\mu\lambda){\cal M}^{-1}\, \del^\mu{\cal M}\Big)\ ,\nn\\
&=& -\ft12 {\rm tr}\, \Big( \del_\mu\lambda({\cal M}^{-1}\, \del^\mu {\cal M} 
+ ({\cal M}^T)^{-1} \del^\mu {\cal M}^T)\Big)\nn\\
&=& -{\rm tr}\, (\del_\mu\lambda \, {\cal M}^{-1}\, \del^\mu{\cal M})\ .
\eea
Thus we obtain the $G$-valued Noether currents
\be
{\cal J} = -{\cal M}^{-1}\, d {\cal M} 
                 \ . \label{noether}
\ee
It is easily verified that under global $G$ transformations
(\ref{globaltrans}), the Noether currents transform as
\be
{\cal J}\longrightarrow {\cal J}' = \Lambda^{-1}\, {\cal J}\, 
\Lambda\ .\label{noethertrans}
\ee

     Applying this to the $SL(3,\R)$ Lagrangian (\ref{sl3rlag}),
described by (\ref{mlag}) with ${\cal M}$ given by (\ref{sl3v}) and
(\ref{mvv}), we find the Lie algebra $SL(3,\R)$-valued Noether currents
\be
{\cal J} = \pmatrix{ {\cal J}_{11} & {\cal J}_{12} & {\cal J}_{13} \cr
                  {\cal J}_{21} & {\cal J}_{22} & {\cal J}_{23}\cr
                 {\cal J}_{31} & {\cal J}_{32} & {\cal J}_{33} }\ ,
\label{sl3rnoether}
\ee
where the components are given by
\bea
{\cal J}_{11} &=&  - \ft1{\sqrt{3}}\, d\phi_1 + d\phi_2 -
   e^{\sqrt{3}\phi_1 - \phi_2}\, \chi_{12}\, d\chi_{12} -
e^{-2\phi_2}\,  \chi_{13}\, \chi_{23}\, d\chi_{12}\nn\\
& &+
   e^{-2\phi_2}\, \chi_{12}\, \chi_{23}^2\, d\chi_{12}+
e^{-2\phi_2}\, \chi_{13}\, d\chi_{13}
 -e^{-2\phi_2}\,  \chi_{12}\,\chi_{23}\, d\chi_{13} \nn \\
{\cal J}_{22} &=&  \ft2{\sqrt{3}}\, d\phi_1 +
e^{\sqrt{3}\phi_1 - \phi_2}\, \chi_{12}\, d\chi_{12}
-e^{-2\phi_2}\, \chi_{12}\, \chi_{23}^2\, d\chi_{12}\nn\\
&& + e^{-2\phi_2}\,  \chi_{12}\,\chi_{23}\, d\chi_{13}
   -e^{-\sqrt{3}\phi_1 - \phi_2}\, \chi_{23}\, d\chi_{23}\nn\\
{\cal J}_{12} &=&  
      -\sqrt{3}\,\chi_{12}\, d\phi_1 + \chi_{12}\,d\phi_2 - d\chi_{12} -
   e^{\sqrt{3}\phi_1 - \phi_2}\,\chi_{12}^2\,d\chi_{12} -
  e^{-2\phi_2}\, \chi_{12}\,\chi_{13}\,\chi_{23}\,d\chi_{12}\nn \\
& &+ e^{-2\phi_2}\, \chi_{12}^2\,\chi_{23}^2\,d\chi_{12} +
e^{-2\phi_2}\, \chi_{12}\,\chi_{13}\, d\chi_{13} -
e^{-2\phi_2} \, \chi_{12}^2 \,\chi_{23}\, d\chi_{13}  \nn\\
& &- e^{-\sqrt{3}\phi_1 - \phi_2}\, \chi_{13}\, d\chi_{23} +
   e^{-\sqrt{3}\phi_1 - \phi_2}\, \chi_{12}\, \chi_{23}\, d\chi_{23}\nn\\
{\cal J}_{13} &=& -\sqrt{3}\, \chi_{12}\, \chi_{23}\, d\phi_1 
     + 2\chi_{13}\, d\phi_2 -
 \chi_{12}\, \chi_{23}\, d\phi_2 -
   e^{\sqrt{3}\phi_1 - \phi_2}\, \chi_{12}\, \chi_{13}\, d\chi_{12}\nn \\
& & - e^{-2\phi_2} \, \chi_{13}^2\, \chi_{23}\, d\chi_{12}  +
 e^{-2\phi_2}\,  \chi_{12}\, \chi_{13}\, \chi_{23}^2\, d\chi_{12}
 -d\chi_{13} \nn\\
&&+e^{-2\phi_2}\, \chi_{13}^2\, d\chi_{13} -
 e^{-2\phi_2} \,\chi_{12}\, \chi_{13}\, \chi_{23}\, d\chi_{13}\nn \\
& & + \chi_{12}\, d\chi_{23} - e^{-\sqrt{3}\phi_1 -
         \phi_2}\, \chi_{13}\, \chi_{23}\, d\chi_{23} +
   e^{-\sqrt{3}\phi_1 - \phi_2}\, \chi_{12}\, \chi_{23}^2\, d\chi_{23} \nn \\
{\cal J}_{21} &=&  e^{\sqrt{3}\phi_1 - \phi_2}\, d\chi_{12} - e^{-2\phi_2}\,
\chi_{23}^2\, d\chi_{12} +e^{-2\phi_2}\,
   \chi_{23}\, d\chi_{13}\nn\\
{\cal J}_{23} &=& \sqrt{3}\chi_{23}d\phi_1 +\chi_{23}\, d\phi_2 +
   e^{\sqrt{3}\phi_1 - \phi_2}\, \chi_{13}\, d\chi_{12} -
      e^{-2\phi_2} \,\chi_{13}\, \chi_{23}^2\, d\chi_{12}\nn \\
 & &+ e^{-2\phi_2}\, \chi_{13}\,\chi_{23}\, d\chi_{13} - d\chi_{23} -
   e^{-\sqrt{3}\phi_1 - \phi_2}\, \chi_{23}^2\, d\chi_{23}\nn\\
{\cal J}_{31} &=& e^{-2\phi_2}\,\chi_{23}\, d\chi_{12} -e^{-2\phi_2}\,
d\chi_{13}\nn\\
{\cal J}_{32} &=& e^{-2\phi_2}\, \chi_{12}\,\chi_{23}\,d\chi_{12}-
e^{-2\phi_2}\,\chi_{12}\,d\chi_{13} +
   e^{-\sqrt{3}\phi_1 - \phi_2}\, d\chi_{23}\nn \\
\eea
and ${\cal J}_{33} = -{\cal J}_{11} -{\cal J}_{22}$.

\end{document}